\documentclass[%
 preprint,
 amsmath,amssymb,
 aip,pop,
 floatfix,
]{revtex4-2}

\usepackage{xcolor}
\usepackage{graphicx}
\usepackage{dcolumn}
\usepackage{bm}

\usepackage[utf8]{inputenc}
\usepackage[T1]{fontenc}
\usepackage{mathptmx}
\usepackage{hyperref}
\begin{document}



\title{
Experimental observations of detached bow shock formation in the interaction of a laser-produced plasma with a magnetized obstacle
}

\author{Joseph M. Levesque}
\email{jmlevesq@umich.edu}
\affiliation{P-2, Fundamental and Applied Physics, Los Alamos National Laboratory, Los Alamos, New Mexico, 87507}
\affiliation{Department of Climate and Space Sciences and Engineering, University of Michigan, Ann Arbor, Michigan 48109}
\affiliation{Department of Nuclear Engineering and Radiological Sciences, University of Michigan, Ann Arbor, Michigan 48109}

\author{Andy S. Liao}
\affiliation{Department of Physics and Astronomy, Rice University, Houston, Texas 77005}
\affiliation{Los Alamos National Laboratory, Los Alamos, New Mexico 87507}

\author{Patrick Hartigan}
\affiliation{Department of Physics and Astronomy, Rice University, Houston, Texas 77005}

\author{Rachel P. Young}
\author{Matthew Trantham}
\author{Sallee Klein}
\affiliation{Department of Climate and Space Sciences and Engineering, University of Michigan, Ann Arbor, Michigan 48109}
\affiliation{Department of Nuclear Engineering and Radiological Sciences, University of Michigan, Ann Arbor, Michigan 48109}

\author{William Gray}
\affiliation{Department of Climate and Space Sciences and Engineering, University of Michigan, Ann Arbor, Michigan 48109}

\author{Mario Manuel}
\affiliation{Department of Climate and Space Sciences and Engineering, University of Michigan, Ann Arbor, Michigan 48109}
\affiliation{General Atomics, San Diego, California 92121}

\author{Gennady Fiksel}
\affiliation{Department of Nuclear Engineering and Radiological Sciences, University of Michigan, Ann Arbor, Michigan 48109}

\author{Joseph Katz}
\affiliation{Laboratory for Laser Energetics, Rochester, New York 14623}

\author{Chikang Li}
\author{Andrew Birkel}
\affiliation{Plasma Science and Fusion Center, Massachusetts Institute of Technology, Cambridge, Massachusetts 02139}

\author{Petros Tzeferacos}
\affiliation{Department of Astronomy and Astrophysics, University of Chicago, Chicago, Illinois 60637}
\affiliation{Department of Physics and Astronomy, University of Rochester, Rochester, NY 14627}
\affiliation{Laboratory for Laser Energetics, Rochester, New York 14623}

\author{Edward C. Hansen}
\affiliation{Department of Astronomy and Astrophysics, University of Chicago, Chicago, Illinois 60637}
\affiliation{Department of Physics and Astronomy, University of Rochester, Rochester, NY 14627}

\author{Benjamin Khiar}
\affiliation{Department of Astronomy and Astrophysics, University of Chicago, Chicago, Illinois 60637}

\author{John M. Foster}
\affiliation{Atomic Weapons Establishment, Aldermaston, Reading RG7 4PR, UK}

\author{Carolyn C. Kuranz}
\affiliation{Department of Climate and Space Sciences and Engineering, University of Michigan, Ann Arbor, Michigan 48109}
\affiliation{Department of Nuclear Engineering and Radiological Sciences, University of Michigan, Ann Arbor, Michigan 48109}

\begin{abstract}


The magnetic field produced by planets with active dynamos, like the Earth, can exert sufficient pressure to oppose supersonic stellar wind plasmas, leading to the formation of a standing bow shock upstream of the magnetopause, or pressure-balance surface.
Scaled laboratory experiments studying the interaction of an inflowing solar wind analog with a strong, external magnetic field are a promising new way to study magnetospheric physics and to complement existing models, although reaching regimes favorable for magnetized shock formation is experimentally challenging.
This paper presents experimental evidence of the formation of a magnetized bow shock  in the interaction of a supersonic, super-Alfv\'enic plasma with a strongly magnetized obstacle at the OMEGA laser facility.
The solar wind analog is generated by the collision and subsequent expansion of two counter-propagating, laser-driven plasma plumes.
The magnetized obstacle is a thin wire, driven with strong electrical currents.
Hydrodynamic simulations using the FLASH code predict the colliding plasma source meets the criteria for bow shock formation.
Spatially resolved, optical Thomson scattering measures the electron number density, and optical emission lines provide a measurement of the plasma temperature, from which we infer the presence of a fast magnetosonic shock far upstream of the obstacle.
Proton images provide a measure of large-scale features in the magnetic field topology, and reconstructed path-integrated magnetic field maps from these images suggest the formation of a bow shock upstream of a the wire and as a transient magnetopause.
We compare features in the reconstructed fields to two-dimensional MHD simulations of the system.

\end{abstract}

\maketitle

\section{Introduction}\label{Sec:Introduction}

Astrophysical plasmas are typically magnetohydrodynamic, and because of the large spatial and temporal scales over which these plasmas evolve, magnetic fields are dynamically important throughout much of the universe.
For example, the magnetic field pressure produced by magnetized obstacles in these plasma flows can separate the obstacle from the surrounding plasma and form a magnetosphere.
Stellar winds are typically supersonic and super-Alfv\'enic, so in the interaction of a plasma wind with strongly magnetized bodies a bow shock forms upstream of the physical obstacle to redirect the flow.\cite{Spreiter_PSS1966}
In planetary magnetosphere systems, the effective obstacle to the flow is not the physical surface of the obstacle, but the surface at which the magnetic field and the surrounding plasma reach pressure balance, known as the magnetopause.
The standing, detached bow shock therefore occurs upstream of the magnetopause, and the region of shocked flow between the bow shock and magnetopause is known as the magnetosheath.
There are numerous examples of astrophysical flows past magnetized obstacles that produce shock waves, most notably the shocks that occur when the solar wind impinges upon a planetary magnetosphere, like for the Earth or Jupiter.
In the Earth-Sun system, the Earth's magnetic field exerts sufficient pressure to oppose the solar wind at an average distance of $\sim$10 Earth radii, with the bow shock $\sim$3 Earth radii farther upstream.\cite{Fairfield_JGR1971}

Semi-analytical and empirical models of the Earth's magnetosphere based on satellite observations can accurately determine the location of the magnetopause and bow shock using the history of observed solar wind parameters.\cite{Spreiter_PSS1966,Fairfield_JGR1971,Zhuang_JGR1981,CairnsGRL1994}
Interest in magnetized bow shocks is not limited to just the Earth system, nor even limited to those within the Solar system; with thousands of exoplanets already discovered, there is growing interest in learning how stellar winds interact with both weakly magnetized and strongly magnetized planets under a range of conditions beyond those normally present in our solar system.
For example, exoplanets that possess little or no intrinsic magnetic fields (like Venus and Mars in our solar system)\cite{Luhmann_Chapter} are more likely than magnetized planets to have their atmospheres stripped away by stellar flare events during the earlier stages of their star’s life, with implications for the atmospheric compositions, densities, and the potential for the development of life on these objects.\cite{Khodachenko_Astrobiology2007,Vidotto_MNRAS2015}
Numerical models of magnetospheric systems are difficult to validate, owing to the large number of free parameters in the system, the need for fine-scale resolution of the shear interfaces simultaneously with large-scale tracing of field geometries in rapidly evolving systems, and the coupling between these scales.\cite{Toth_JGR2005}
However, through proper scaling of system parameters it is also possible to explore physics relevant to large-scale astrophysical phenomena on laboratory scales.\cite{Ryutov_ApJS2000,Ryutov_PoP2001,Ryutov_PPCF2002}
A laboratory analog capable of testing the interaction of variable plasma flows with magnetized obstacles would complement the numerical models.
Tunable experimental magnetosphere systems could provide new insights into, for example, the early stages of bow shock formation, how the magnetosphere responds to more extreme plasma conditions\cite{Lavraud2008}, and the role of turbulence and instabilities on the Earth's bow shock and magnetopause.\cite{Moore_NPhys2016}

In this paper we report experimental observations of detached bow shock formation in the interaction of a supersonic, super-Alfv\'enic plasma with a strongly magnetized obstacle at the OMEGA laser facility.
In our experiments the magnetized obstacle is a current-carrying wire. 
We take a somewhat novel approach to create our solar wind analog; we use high-intensity lasers to generate two counter-propagating carbon plasma plumes which collide on-axis and expand outward toward the wire, reaching favorable conditions for magnetized shock formation.
Although the field topology is not dipolar like the Earth's, the field around the wire acts as a suitable analog to a planetary magnetic field for the purposes of generating a shock.\cite{Liao_HEDP2015}
We infer plasma parameters around the location of expected shock formation from spatially resolved, optical Thomson scattering spectra measured with the Imaging Thomson Scattering (ITS) diagnostic.\cite{FroulaBook,Froula_RSI2006,Katz_JInst2013}
We probe the magnetic fields within the system using proton imaging\cite{Li_RSI2006,Li_PoP2009} from which we estimate path-integrated magnetic fields using a new multi-image reconstruction technique, and infer the presence of a bow shock and a transient magnetopause based on the compression of the field.\cite{Levesque_RSI2021}

The primary goal of our experiment was to scale the magnetosphere system down to laboratory scale, to create a magnetopause and measure a bow shock at a significant standoff distance from the physical obstacle, driven by the magnetic field pressure opposing the incoming plasma.
Therefore, the key consideration for the viability of an experimental platform is the balance between the ram pressure of the incoming plasma flow
\begin{equation}
	P_{\mathrm{ram}} = \rho u^2
\end{equation}
and the magnetic pressure of the externally applied field (in SI units)
\begin{equation}
	P_B = B_{\perp}^2 / 2 \mu_0,
\end{equation}
where $\rho$ and $u$ are the density and velocity of the incoming plasma, $B_{\perp}$ is the magnetic field amplitude aligned perpendicular to the plasma flow, and $\mu_0$ is the vacuum permeability.
For simplicity, we define the ratio of ram to magnetic pressure as the dimensionless parameter 
\begin{equation}
	\beta_{\mathrm{ram}} = P_{\mathrm{ram}}/P_{B} = 2 \rho u^2 \mu_0 / B_{\perp}^2,
\end{equation}
and the condition for the formation of a magnetopause is $ \beta_{\mathrm{ram}} \leq 1$.
In order for a bow shock to form upstream of this magnetopause, the plasma velocity must also exceed the signal speed of the plasma, which, in the case of a magnetic field aligned perpendicular to the direction of plasma flow, is the fast magnetosonic speed
\begin{equation}\label{eq:vfms}
	v_{fms} = \sqrt{v_A^2+c_s^2},
\end{equation}
where $v_A$ is the  Alfv\'en speed and $c_s$ is the sound speed.
In addition to $\beta_{ram}$ and $v_{fms}$, there are three other constraints to consider: (i) the timescales must be long enough for a shock to form, (ii) magnetic Reynolds numbers must be large enough to ensure the field does not quickly diffuse away during the experiment, and (iii) the interaction length of the incoming plasma must be smaller than the size of the obstacle to form a shock.
In a collisionless shock like the Earth's bow shock the gyroradius sets the interaction length, whereas in our experiment the interaction length is defined by a collisional mean free path. 
For the purposes of creating a magnetized shock it doesn't matter which operates as long as the the plasma interaction length scale is smaller than the obstacle.

Ours are not the only experiments pursuing laboratory-scale magnetized shocks, recent z-pinch experiments at the MAGPIE facility\cite{Burdiak_PoP2017,Suttle_2020PPCF,Suttle_RSI2021} have demonstrated shock formation in the interaction of a magnetized plasma with initially unmagnetized, conducting obstacles, as well as in the interaction of a magnetized plasma with an external, magnetized obstacle at laboratory scales.
And, more similar to our experiments, \citet{Rigby_NatPhys2018} provide evidence of magnetized shock formation in the interaction of a laser-produced plasma with a magnetized sphere.
However, unlike these other experiments, proton imaging allows us to probe the magnetic field of the system over a large field of view, and we observe the evolution of the system by obtaining a time series of proton images over multiple shots.

Section \ref{Sec:ExperimentalSetup} further describes the experimental setup --- the magnetized obstacle, the colliding plasma flow source and its advantages in comparison to the plasma from a single laser-irradiated target, and the relative orientation of the diagnostics.
Section \ref{Sec:ThomsonScattering} presents the inferred electron number density and electron temperature from ITS measurements of what we infer to be a shock transition at a significant standoff distance from the wire.
Section \ref{Sec:ProtonImaging} presents the proton images of the experiment for two applied field strengths, and by using field reconstruction we infer the formation of a bow shock upstream of the magnetized wire obstacle.
Section \ref{Sec:MHDSimulations} presents results of two-dimensional MHD simulations of the experiment and compares these simulations to the data.
Section \ref{Sec:Discussion} further discusses some important aspects of the achieved parameters, and addresses the experiment's similarity to the Earth's magnetosphere.
Finally, section \ref{Sec:Conclusion} concludes the paper and presents some considerations for future experiments using this platform.

\begin{figure*}[ht!]
	\centering
	\includegraphics{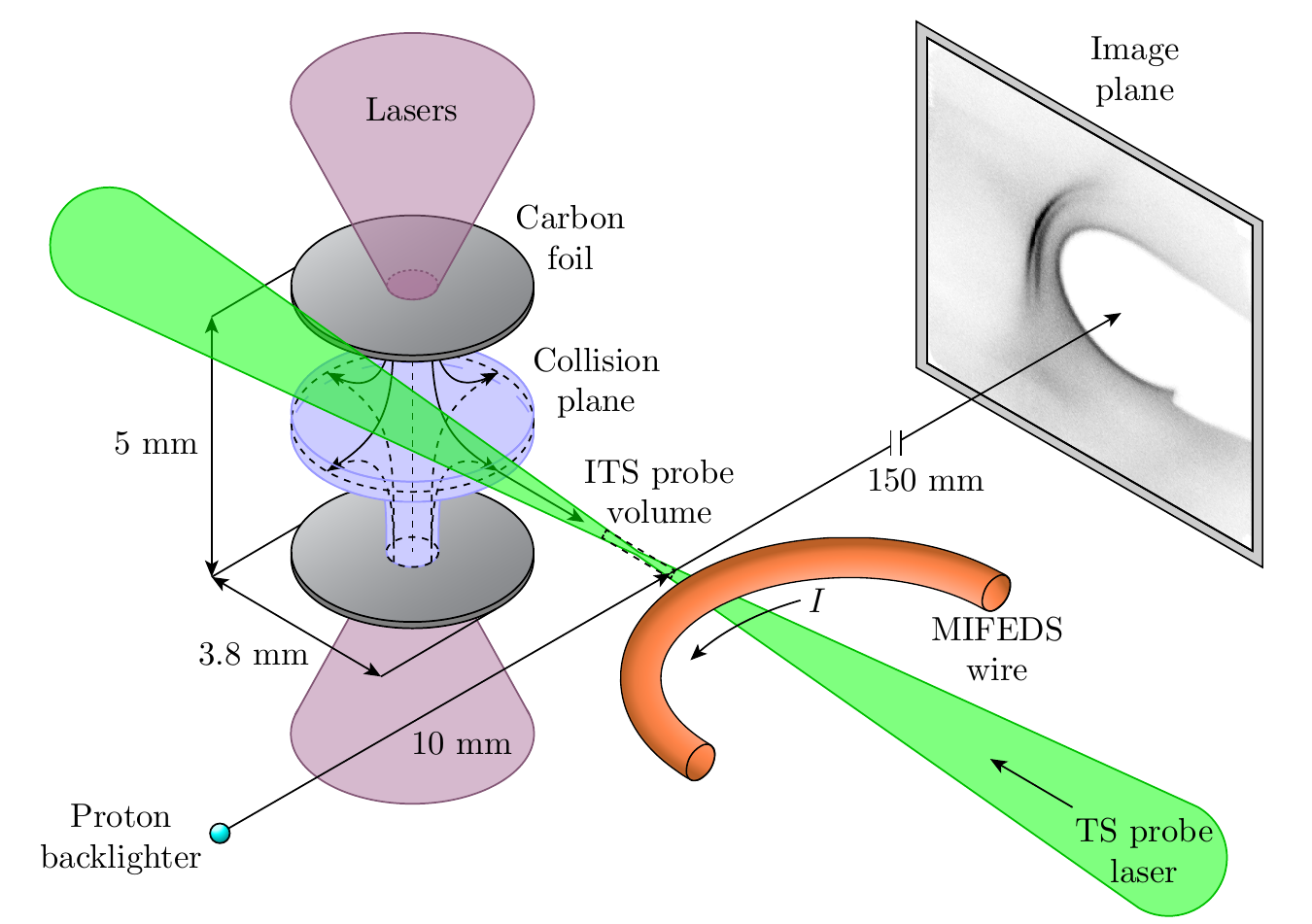}
	\caption{
		Illustration of the experiment with some key length scales given.
		Two thin carbon discs are simultaneously irradiated with six 450 J, 351 nm lasers over a 1 ns square pulse within a 800 $\mu$m spot size, to generate counter-propagating plasma flows.
		The flows collide and expand outward toward the magnetized obstacle, which is a thin current-carrying wire driven via the MIFEDS.
		This design produces a steadier and lower-density flow than that generated from a single laser-irradiated disc, which therefore has more time to create a magnetized bow shock before the densest part of the flow overruns the wire.
		The ITS diagnostic measures the spectrum of Thomson scattered light from a 2$ \omega $ probe beam incident at $ 43.7^{\circ} $ to the primary flow axis 1.45 mm upstream of the wire in a 1.8 mm field of view along the laser axis.
		Protons produced by the implosion of a D$^3$He capsule probe the magnetic fields in the area of interest, and are captured by a CR-39 detector 16 cm away (proton image not to scale).
	}
	\label{Fig:Setup}
\end{figure*}

\section{Experimental Setup} \label{Sec:ExperimentalSetup}

\begin{figure*}[t]
	\centering
	\includegraphics{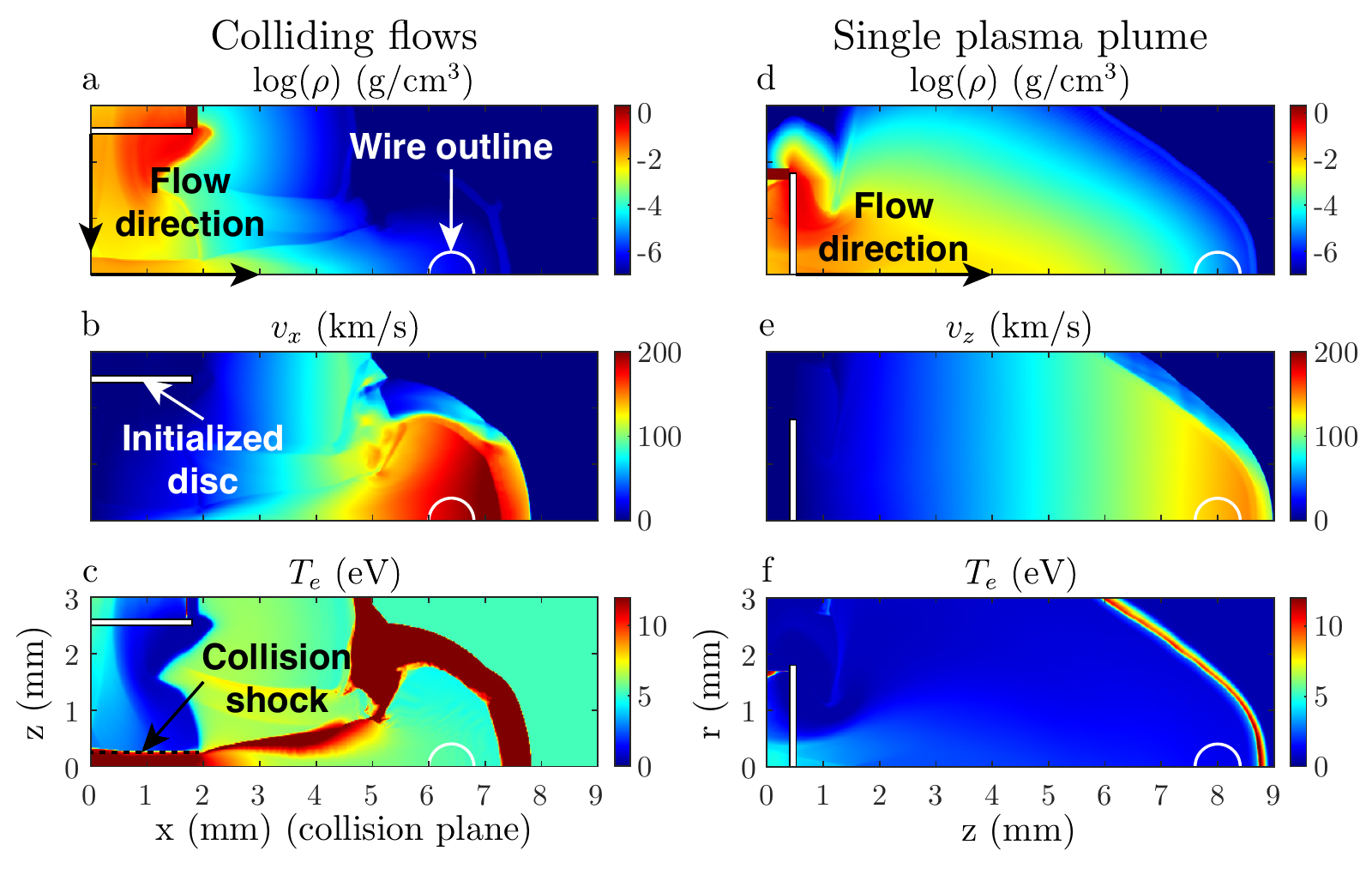}
	\caption{
	Results from FLASH hydrodynamic simulations of the colliding flow source and a single plasma plume, 60 ns after laser drive, for comparison, with some key features labeled.
	(a--c) The density, velocity in the $x$ direction, and electron temperature for the colliding flow source.
	(d--f) The density, velocity in the $z$ direction, and electron temperature for the single plasma plume.
	Each simulation is cylindrically symmetric about its $z$ axis, along which the laser is incident, and the plane $z=0$ is a reflecting boundary for the colliding flows.
	The thin, white boxes at the left of each image show the initial extent of the solid carbon discs for each simulation.
	The white semi-circles in the colliding flow images outline the nominal location of the wire in the experiment, and for the single foil case a wire at a corresponding axial distance from the carbon foil would have a forward surface at $z=9$ mm (this obstacle was not included in these simulations).
	}
	\label{Fig:SourceComp}
\end{figure*}

The experiment is illustrated in Figure \ref{Fig:Setup}, for which the primary components are the magnetized obstacle and the laser-irradiated source of plasma inflow.
The magnetized obstacle is created by using the OMEGA Magneto-Inertial Fusion Electrical Discharge System (MIFEDS)\cite{Fiksel_RSI2015} to drive up to 25.5 kA of current in a 760 $\mu$m diameter, kapton-insulated copper wire.
The nominal 4 mm radius of curvature of the wire with respect to the central experimental plane of Figure \ref{Fig:Setup} is large compared to the system scale, determined by the wire radius, so the field can be approximated fairly well as the field around an infinite current-carrying wire, which drops off as $R/r$ where $R$ is the radius of the wire and $r$ is radial distance.
To compare the effect of different magnetic field strengths, two nominal driven currents in the wire were used --- 25.5 kA and 17 kA --- which generate maximum nominal magnetic fields at the surface of the wire of 13.5 T and 9 T, respectively. 
We refer to the two nominal currents as the high-field and low-field configurations.
Despite the relatively large maximum magnetic fields achieved by MIFEDS at the surface of the wire, reaching the desired parameter space at a measurable standoff distance requires tailoring the plasma flow source to suitable densities and velocities.
For reference: for a current-carrying wire with a maximum surface field of 13.5 T, at a radial distance of 1.45 mm from the wire the field decreases to roughly 3.5 T and the magnetic pressure there is predicted to oppose a plasma with density $10^{-6}$ g/cm$^3$ traveling at 100 km/s, assuming no dynamic compression of the field. 

Laser-generated plasma sources are typically highly energetic, so a novel approach was used to achieve the desired parameters.
The plasma source is created by colliding two counter-propagating plasma plumes from two laser-irradiated carbon discs. 
The discs are nominally 3.8 mm in diameter and 100 $\mu$m thick, oriented normal to one another and spaced 5 mm apart along the normal axis as shown in Figure \ref{Fig:Setup}.
The carbon discs are simultaneously irradiated by six 450 J, 351 nm lasers over a 1 ns square pulse within a 800 $\mu$m spot size (total irradiance $\approx 1.34\times 10^{14}$ W/cm$^2$), driving a shock through each disc.
The plasma flows expand from the surfaces opposite the lasers, collide on-axis at the midplane between the discs, and then expand outward toward the wire, which is 6.45 mm away from the collision axis along the collision plane.

Figure \ref{Fig:SourceComp} shows density, temperature, and velocity results from 2D hydrodynamic (no magnetic field) simulations of the laser-driven colliding plasma source in comparison to a single, directed plasma plume, and demonstrates the utility of this source.
These simulations were performed using the FLASH code, a modular, open-source, Eulerian magnetohydrodynamics code which includes laser energy deposition, multigroup radiation diffusion, and tabulated equations of state (EOS) and opacities.\cite{Lee_JCompPhys2009,Tzeferacos_HEDP2012}
These simulations have 20 $\mu$m resolution, and are cylindrically symmetric about the $z$ axis in each system.
The colliding flow is approximated by colliding a plasma plume against a reflecting boundary condition at $z=0$.
The laser parameters used in simulation mirror that of the experiment, as described above.
It is immediately apparent that the two plasma flows exhibit very different properties.
The expansion of the colliding source reduces the density of the flow that interacts with the wire, and the longer effective path length to the wire reduces the density and velocity gradients when compared to a direct plasma plume.
The reduced density makes shock formation possible, and the flattened gradients extends the duration over which favorable conditions exist.

\begin{figure*}
	\centering
	\includegraphics{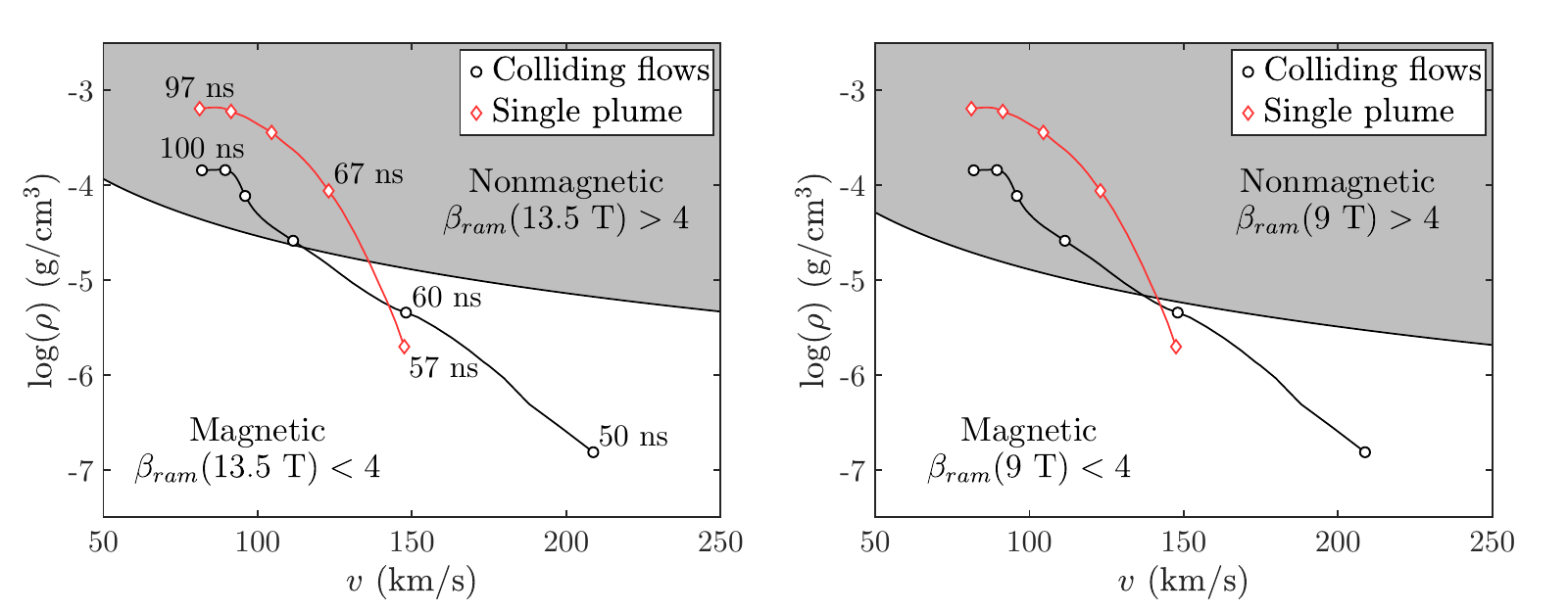}
	\caption{
	Region plots comparing the plasma parameters of the two sources in $\rho$-$v$ space for the high-field case (left), and the low-field case (right).
	The plotted parameters are measured 7.5 mm from the target surface in each case, where the path length in the colliding flow case is first along the axial direction and then the radial direction after colliding. 
	The points begin when the plasma first reaches the point of interest, with the circle and diamond markers at 10 ns intervals.
	The shaded regions show where $\beta_{ram}>4$ (the value 4 corresponds to a doubling of the nominal field at the magnetopause) and the plasma becomes nonmagnetic.
	}
	\label{Fig:RegionPlot}
\end{figure*}

Figure \ref{Fig:RegionPlot} plots the general time evolution of plasma parameters in $\rho-v$ space for the two simulated hydrodynamic flows, measured 1.45 mm upstream from the wire.
The plasma is considered to be magnetized when $\beta_{ram}<4$, and unmagnetized when $\beta_{ram}>4$, and the limits of magnetopause formation for both field configurations are displayed as the non-shaded regions.
The choice of 4 comes from the assumption that the post-shock field at the magnetopause is compressed to $2\times$ the nominal field value.
In an ideal system with a perfectly conducting, uniform plasma inflow, the doubling of the field at the magnetopause arises analytically by placing an image current at twice the distance from the magnetopause, following the early work in magnetospheric physics by \citet{Chapman_Ferraro_1931b}.
Of course, the plasma of this experiment has finite conductivity and is highly dynamic, so the compression is expected to vary in time.
For the colliding flow source, FLASH predicts a density of $\sim 10^{-6}$ g/cm$^3$ and a velocity of $\approx 200$ km/s 50 ns after laser drive, which decreases to $\approx 150$ km/s during the following 10 ns.
Under these conditions, the formation of a bow shock assuming 2$\times$ field compression is predicted to be possible between 50 and 70 ns in the high-field configuration, for up to 3 wire-crossing times, and between 50 and 60 ns in the low-field case, for only about 1 wire-crossing time.
In contrast, the density of the plasma plume produced from a single disc, irradiated with the same laser intensity, is at least an order of magnitude greater than in the colliding flows by the time it reaches the same effective distance. 
The single plasma plume would therefore quickly outstrip the magnetic pressure less than 10 ns after first reaching the wire in either field configuration --- not sufficient time for a magnetopause to develop.
Note that these estimates are based on static assumptions, and the work done by the plasma to compress the field may reduce the velocity below these purely hydrodynamic predictions.

We probe the system using two primary diagnostics: spatially resolved, optical Thomson scattering spectrometry, and proton imaging.
The spatially resolved Imaging Thomson Scattering diagnostic (ITS)\cite{Ross_RSI2006,Katz_JInst2013} measures the Thomson-scattered electron plasma wave (EPW) spectra from the 2$\omega$ (526.5 nm wavelength) probe laser within a 100 $\mu$m $\times$ 1.8 mm cylindrical probe volume.
The TS probe beam is focused 1.45 mm upstream from the wire, where the bow shock was expected to form 60 ns after irradiating the carbon targets.
The probe beam has a $ 43.7^{\circ} $ angle of incidence with respect to the collision flow axis.
Proton imaging uses the laser-driven implosion of a $400$ $\mu$m D${}^3$He capsule as the proton source, which produces quasi-monoenergetic protons at 3 MeV and 14.7 MeV with a $\sim$100 $\mu$m source size.\cite{Li_RSI2006,Li_PoP2009}
The capsule is positioned 1 mm upstream from the wire along the plasma flow axis and 1 cm from the experimental plane.
The protons are captured by a roughly 10 cm square CR-39 detector\cite{Li_RSI2006} placed 15 cm from the experimental plane on the opposite side of the wire from the source.
With this geometry the proton images have a magnification of 16 between the plane at the center of the wire and the image plane.

\section{Thomson Scattering} \label{Sec:ThomsonScattering}

\begin{figure*}[t]
	\centering
	\includegraphics{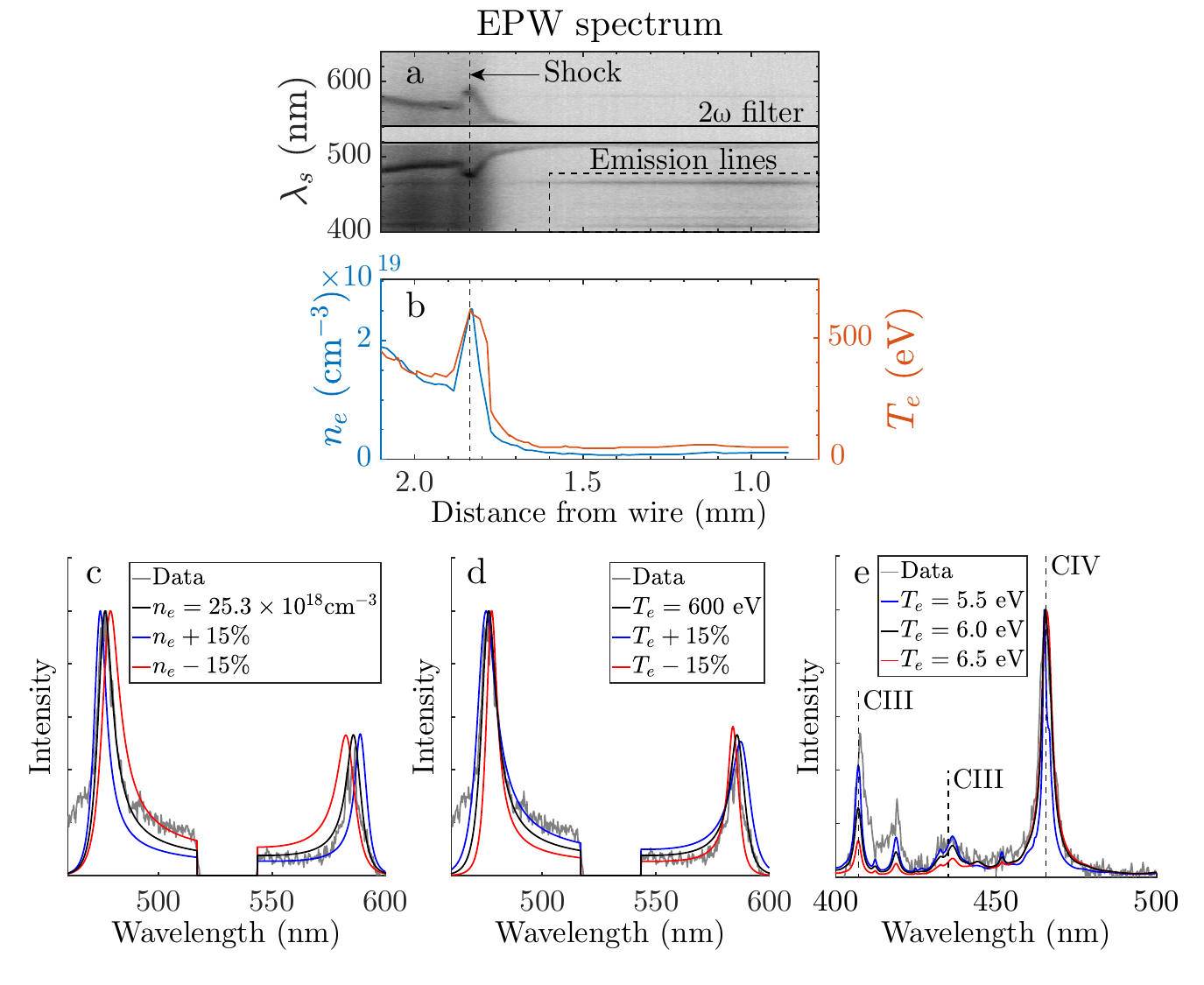}
	\caption{
		(a) The Thomson-scattered electron plasma wave spectrum, measured in wavelength $\lambda_s$, 50 ns after initial laser drive.
		(b) Inferred fits to electron number density and electron temperature from the spectra of (a). 
		The position axis is calibrated relative to the axis from the collision plane to the wire. 
		The inferred shock location is overlaid as a dashed line in (a) and (b).
		(c, d) Best fits to the spectrum at the shock location, demonstrating that plasma parameter fits are typically accurate to within $\pm$ 15 \%, in $n_e$ (c), and $T_e$ (d). 
		(e) Best fit to the self-emission spectrum region outlined in (a) using PrismSPECT and assuming a carbon plasma, the primary peak at 470 nm is from CIV--CIII transitions, while the smaller peaks are CIII--CII transitions.
	}
	\label{Fig:ITS}
\end{figure*}

Thomson scattering of laser light from a plasma is a well-known method of measuring plasma parameters, where, by fitting synthetic Thomson scattered spectra to the measured spectra, we can infer the electron number density and temperature of the plasma.\cite{Froula_RSI2006,FroulaBook,Ross_RSI2010,Follett_RSI2016}
In a measurement 50 ns after the initial laser drive, using the high-field setup with a 100 ps, 20 J, 526 nm wavelength probe beam, there is a sharp shift of the peak wavelength of the Thomson scattered electron plasma wave spectrum and a corresponding increase in background brightness 1.85 mm upstream from the wire, as shown in Figure \ref{Fig:ITS}(a).
The off-axis incidence of the probe beam complicates the orientation of the spatial axis, so we orient our results based on distance from the wire with respect to the primary axis of the incoming plasma flow.
Based on the dynamics of the nominal plasma flow from simulations, we expect that over the short TS probe the entire system is effectively static.
If we consider the hydrodynamics of just the incoming flow moving at 150 $\mu$m/ns, then over the 100 ps probe duration, the features would travel at most 15 $\mu$m, which is roughly $1\%$ of the probe window, or only a few pixels.

The best fits to $n_e$ and $T_e$ are plotted in Figure \ref{Fig:ITS}(b), with errors of $\le$15\% in the fits, as shown in Figure \ref{Fig:ITS}(c) and \ref{Fig:ITS}(d) for the spectra taken at the location of maximum shift in the scattered wavelength.
We infer an approximate doubling of the electron number density from approximately $12 \times 10^{18} \text{ cm}^{-3}$ in the incoming flow to $25 \times 10^{18} \text{ cm}^{-3}$ over a roughly 100 $\mu$m region between 1.9 and 1.8 mm from the wire along the detector line of sight.
The number density then decreases sharply moving downstream from the spike and toward the wire, to less than $3 \times 10^{18} \text{ cm}^{-3}$ at positions closer than 1.6 mm from the wire, where we can only resolve at most one peak outside of the $2\omega$ filter.
This rapid change in number density from the incoming flow is indicative of a shock, and the factor of 2 increase is an important indicator of the system parameters. 
The large $>100$ eV inferred temperature is likely a result of heating by the intense probe laser, which is expected to primarily occur over a few $10$s of ps.
We use the PrismSPECT software\cite{PrismSPECT_v6.5.0} to calculate synthetic self-emission lines of a hot carbon plasma, and compare the resulting spectra to the self-emission lines (labeled in Figure \ref{Fig:ITS}(a)) that we observe alongside the Thomson scattered spectrum, away from the scattering peaks. 
We infer a best fit temperature of $\sim6$ eV, shown in Figure \ref{Fig:ITS}(e), which corresponds to the temperature of the plasma prior to probe heating and agrees with the temperature predicted by the hydrodynamic simulation of the colliding flows shown in Figure \ref{Fig:SourceComp}(c).

\begin{figure}
	\centering
	\includegraphics{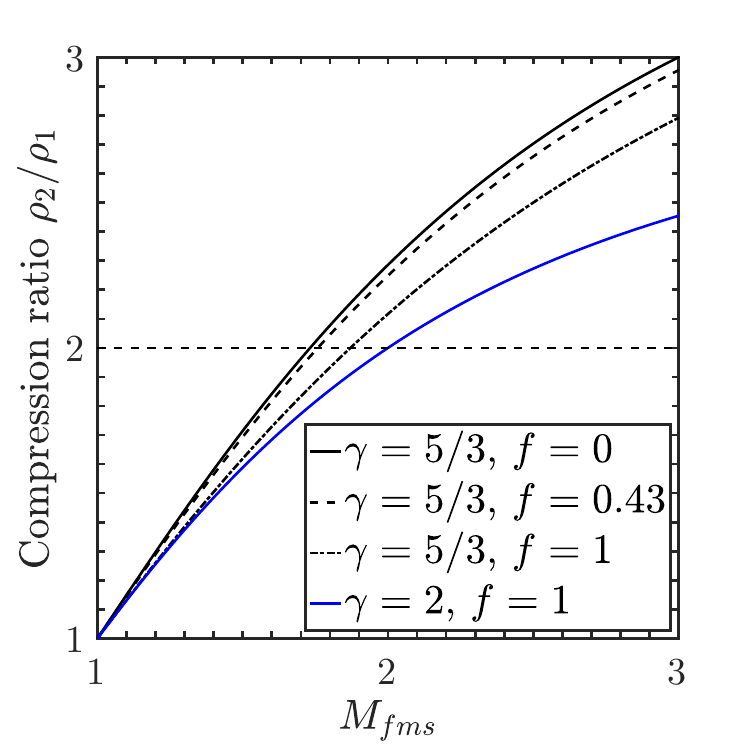}
	\caption{
	Plots of the shock compression ratio as a function of fast magnetosonic Mach number for a perpendicular MHD shock.
	Multiple values of $f$ --- the ratio of Alfv\'en speed to fast magnetosonic speed --- are shown at $\gamma=5/3$.
	A plot for a fully MHD shock at $\gamma=2$ is also shown.
	The dashed line at a compression of 2 corresponds to the observed compression inferred from the Thomson scattered spectra at 50 ns.
	}
	\label{Fig:Compression}
\end{figure}

\begin{table*}[t]
\begin{ruledtabular}
	\centering
	\begin{tabular}{c c c c c c c c}
	$\gamma$ & $f$ & $M_{fms}$ & $T$ (eV) & $c_s$ (km/s) & $B_0$ (T) & $v_{A}$ (km/s) & $v$ (km/s) \\
	\colrule
	5/3 & 0 & 1.73 & 350 & 167 & 0 & -- & 289 \\
	5/3 & 0 & 1.73 & 6 & 15.5 & 0 & -- & 31 \\
	5/3 & 0.43 & 1.76 & 350 & 167 & 2.85 & 80 & 326 \\
	5/3 & $\approx$1 & 1.87 & 6 & 15.5 & 2.85 & 80 & 152 \\
	2 & $\approx$1 & 2 & 6 & 17 & 2.85 & 80 & 164 \\
	\end{tabular}
	\end{ruledtabular}
	\caption{
	Sets of parameters for a perpendicular, fast magnetosonic shock which result in the observed compression of 2 at a distance of 1.8 mm from the wire.
	The 350 eV temperature corresponds to the probe-heated plasma temperature, while the 6 eV temperature is the temperature predicted by simulation and inferred from self-emission spectra.
	The 2.85 T magnetic field is the amplitude of the unperturbed field at a radius of 1.8 mm.
	From these parameters we infer that the shock must be highly magnetized.
	}
	\label{Tab:Compression}
\end{table*}

We next consider what parameters would produce the observed compression ratio of 2.
We assume that the shock is a product of steady flow to simplify analysis, but we can see from Figure \ref{Fig:ITS}a that the density upstream of the shock changes in time.
With an estimated $\sim150$ km/s flow velocity, the upstream density is expected to be nearly uniform for approximately 1 ns after the measurement of Figure \ref{Fig:ITS}a before increasing, so our steady assumption should hold fairly well.
The shock jump conditions for magnetized plasmas are significantly more complex than for their nonmagnetic counterparts, and depend on the angle $\theta$ between the magnetic field and the direction normal to the plane of the shock.\cite{PriestBook,Draine_AnnuRevAA1993,Hartigan2003}
However, the equations simplify when $\theta=90^{\circ}$, so the magnetic field is perpendicular to the shock normal as is the case in this experiment. 
In this limit, the relevant signal speed is the fast magnetosonic speed defined in equation (\ref{eq:vfms}).

Reproducing equations (5) and (6) of \citeauthor{Hartigan2003} (\citeyear{Hartigan2003}), the ratio of the post-shock density $\rho_2$ to the pre-shock density $\rho_1$ for a perpendicular shock in an ideal MHD system is given by 
\begin{equation}
	\frac{\rho_2}{\rho_1} = \frac{2(\gamma+1)}{D+\sqrt{D^2+4(\gamma+1)(2-\gamma)M_A^{-2}}},
\end{equation}
in which $D$ is the quantity
\begin{equation}
	D = \gamma-1+\frac{2}{M^2}+\frac{\gamma}{M_A^2},
\end{equation}
$\gamma$ is the polytropic index of the plasma, $M$ is the Mach number, and $M_A$ is the Alfv\'enic Mach number.
We use these equations to plot in Figure \ref{Fig:Compression} the density jump conditions for a perpendicular MHD shock as a function of the fast magnetosonic Mach number $M_{fms} = v/v_{fms}$ of the incident flow.
Curves are plotted for multiple values of $f=M_A/M_{fms}$ --- the ratio of Alfv\'en speeds to fast magnetosonic speed --- at $\gamma=5/3$, and for a fully MHD shock at $\gamma=2$.
Although the plot shows that the necessary $M_{fms}$ for a compression of 2 does not vary significantly --- between 1.7 and 2 --- for changing $f$ or $\gamma$ it is important to consider what this ratio means for the underlying system parameters.
Table \ref{Tab:Compression} shows possible sets of system parameters which produce a compression of 2 across a perpendicular shock.
Importantly, the velocities $v$ calculated for each set of parameters at the necessary $M_{fms}$ in Table \ref{Tab:Compression} allow us to determine which system is most like the experiment.

First we consider the conditions if the system were completely unmagnetized --- $f=0$ and the shock would is purely hydrodynamic --- the Mach number which produces a doubling of the density in this case is $M=1.73$.
If we assume the shock is due to plasma conditions after probe heating by the Thomson scattering laser, the corresponding flow velocity would be 289 km/s, which is much faster than we predict for the system at this time.
If we instead use the inferred pre-probe temperature of 6 eV, the plasma velocity would be only 31 km/s, which is much slower than predicted.
The parameters necessary for a hydrodynamic shock to provide the observed jump are thus well outside of what we expect for the system at 50 ns, either from simulations (Figure \ref{Fig:SourceComp}) or intuition (distance traveled by the plasma divided by time).

Next we include the magnetic field, assuming a pre-shock field amplitude of 2.85 T perpendicular to the flow, corresponding to a radius of 1.8 mm from the wire in the high-field case.
The Alfv\'en velocity of 80 km/s is significant in comparison with the sound speed.
Using the probe-heated plasma temperature, $f=0.43$ and the plasma velocity would be 326 km/s, more than $100$ km/s faster than we predict for this system.
At 6 eV, however, $f=0.98\approx1$, so the system can be considered completely magnetized, and the velocities of $152$ km/s and $164$ km/s for $\gamma=5/3$ and $\gamma=2$, respectively, are well within the range predicted by FLASH between 50 and 60 ns.

Implicit in the above calculations is the assumption that the plasma source is interacting with a negligible background plasma or vacuum.
There is, however, expected to be some minimal outflow produced from the surface of the current-carrying wire as it undergoes resistive heating from the high current.
In order for any outflow to stop or significantly impede the incoming plasma it would need to have a ram pressure approximately equal to that of the incoming flow.
\citeauthor{Gotchev_RSI2009} (\citeyear{Gotchev_RSI2009}) measured surface temperatures exceeding 5000 K for similar MIFEDS-driven wire conditions, and if we assume that the heated material expands outward at a thermal velocity of $\sim$2 km/s, this outflow would need to have a density in excess of $1$ mg/cm$^{-3}$ --- at least 1000 times greater than that of the incoming flow --- to be significant.
At 50 ns the inferred electron number density from Thomson scattering is $<1 \times 10^{18}$ cm$^{-3}$, which corresponds to a density of $<6 \times 10^{-6}$ g/cm$^{-3}$ for 3 times ionized carbon.
This provides an upper bound on the density of the outflow, which should therefore be negligible.

Thus, using the density measured by Thomson scattering and the simulated velocities for this system, a strongly magnetized plasma is necessary to achieve the relatively low shock compression ratio of 2 that we infer from the Thomson scattered spectra, as a purely hydrodynamic shock would be expected to have a compression ratio of 4.
Under these conditions, the shock appears to be a fast magnetosonic shock, similar to that exhibited by the Earth's bow shock. 
Despite observing a shock, we cannot directly infer how fast the shock is moving in the frame of the flow, or whether it is steady.
Additionally, ITS does not measure the field evolution, so we cannot directly determine whether the field is holding back the flow from this data alone.
Fortunately, the proton images presented in the next section complement the ITS data by probing the magnetic field topology of the system, allowing qualitative and some quantitative analysis of the field structure.
We measured additional Thomson scattering spectra alongside the proton imaging shots, but because of significant changes in the probe beam for those shots --- set to $\sim 300$ J over a 1 ns pulse --- and some measurement issues, we do not include them in this paper.
The additional ITS measurements may be found in the Supplementary Material file.

\section{Proton Imaging} \label{Sec:ProtonImaging}

\begin{figure*}
	\centering
	\includegraphics{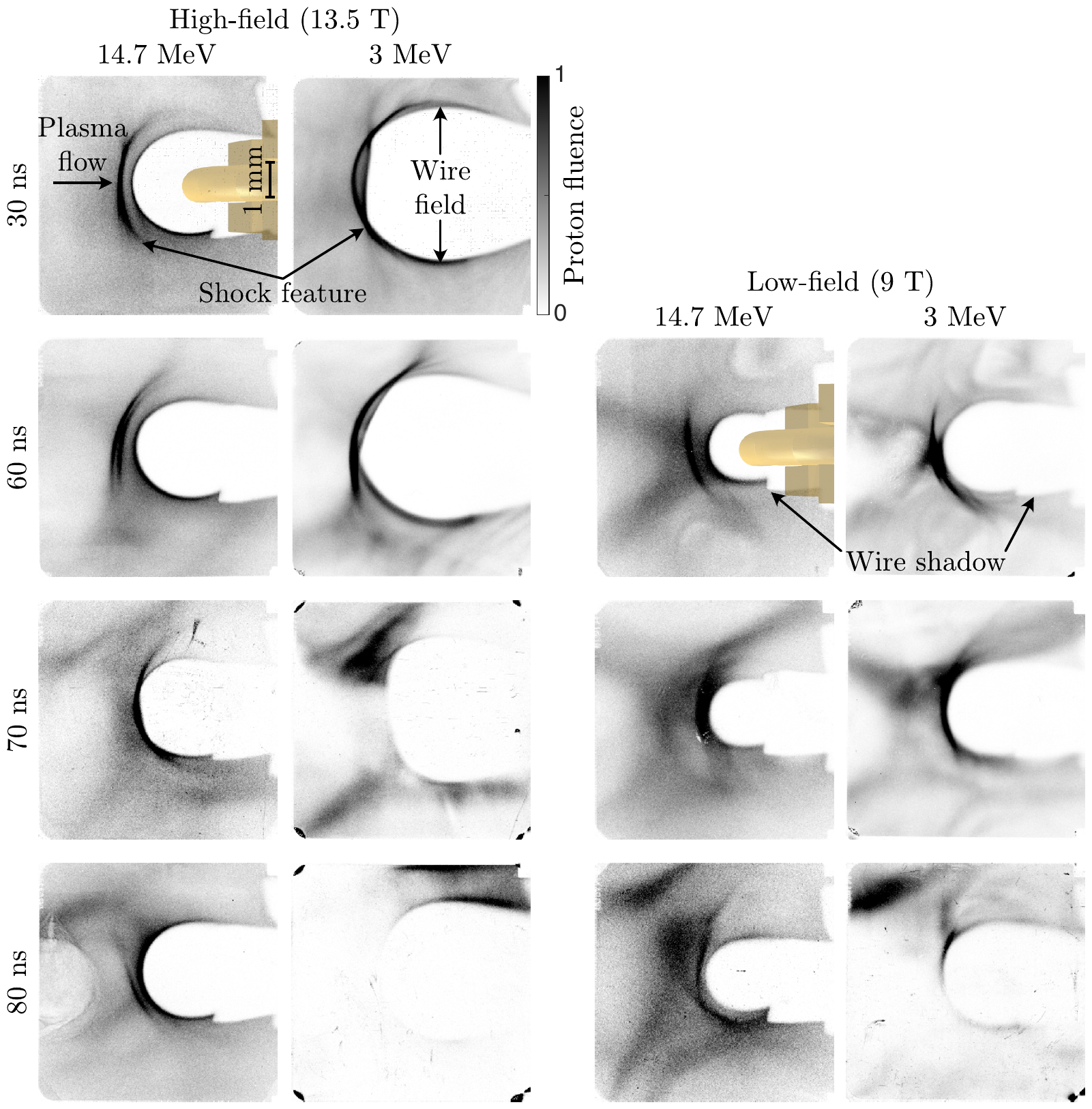}
	\caption{
	Series of 14.7 MeV and 3 MeV proton images from the experiment at various times after source initialization for both high-field and low-field configurations.
	The images are normalized to the highest proton fluence in each image, and the lowest and highest 1\% of values have been saturated to increase contrast of the large-scale features.
	Darker regions indicate increased proton fluence.
	There are four primary features on these images: the wire shadow caused by the wire and wire mount, the caustic wire field feature caused by the nominal field generated by the wire, the shock feature caused by magnetic field compression, and the plasma flow feature corresponding to the inflow.
	A 3D representation of the wire target has been overlaid to demonstrate the wire shadow.
	The image scale corresponds to object plane distances.
	}
	\label{Fig:ProtonImages}
\end{figure*}

\begin{figure*}[t]
	\centering
	\includegraphics{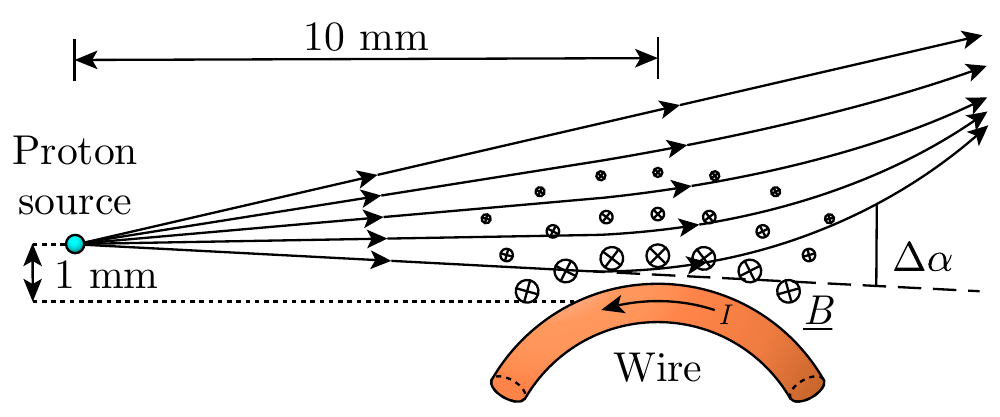}
	\caption{
	Illustration of proton deflection by the field around a current-carrying wire.
	The protons travel primarily antiparallel to the direction of current and are deflected radially outward from the wire.
	Protons that travel closer to the wire experience a stronger magnetic field and undergo larger deflections.
	A caustic forms at the image plane where proton trajectories cross, creating the wire field feature.
	}
	\label{Fig:DeflectionIllustration}
\end{figure*}

Proton imaging measures path-integrated electromagnetic fields based on the deflected positions of high-energy probe protons at the image plane.\cite{Kugland_RSI2012}
In the case of purely magnetic fields, the probe protons are deflected based on the Lorentz force ($\underline{v} \times \underline{B}$) by fields transverse to a proton's trajectory.
Images are formed based on the incident fluence of the 3 MeV and 14.7 MeV protons captured on a 10 cm square CR-39 detector.
Because of the high proton energy, deflections are primarily dependent on the magnetic field's topology in relation to a probe proton's initial trajectory. 
Proton imaging is therefore particularly useful for probing field gradients, but requires a thorough analysis of the system geometry to understand and predict image features.

Figure \ref{Fig:ProtonImages} shows the 14.7 MeV and 3 MeV proton images obtained from this campaign for both field configurations at 30, 60, 70, and 80 ns.
There are many features in these images, so we will first describe the known components.
First, the labeled ``Wire shadow'' feature is the shadow of the solid wire and the structure holding it, which strongly scatter the protons; a 3D representation of this structure has been overlaid on a few images to demonstrate its extent.
The shadow is present to varying degrees in all images, and is characterized by a sharp cutoff in proton fluence without a corresponding increase outside of the shadow.
Next, the ``Wire field'' feature is the (mostly) annular increase in proton fluence away from the wire.
To understand this feature's origin, consider the geometry of the magnetic field with respect to the primary probe axis.
The protons probe antiparallel to the direction of the current in the wire, causing the protons to be deflected radially outward from the wire, as illustrated in Figure \ref{Fig:DeflectionIllustration}. 
Were the direction of the current relative to the probe direction reversed, the protons would instead be radially focused to a point.
The outward deflection of protons simplifies analysis by spreading field information out over a larger area on the image, where in the reverse configuration information is more easily lost by overlapping proton trajectories.

The magnetic field is inversely proportional to distance from the wire, so protons that pass closer to the wire undergo larger deflections.
By assuming a diverging, point source of protons, the proton trajectories that interact with the field farther from the wire will experience smaller deflections and remain mostly ballistic.
A region of increased proton fluence --- a caustic, in this case --- occurs on the image at the point where the highly deflected trajectories and the more ballistic trajectories of the protons cross, resulting in the wire field feature.
The region closer in to the wire from this crossing point is devoid of protons, as they have been deflected farther out than the intensity maximum.
For even modest field amplitudes, this combination of probe orientation and field topology will produce a caustic, making analysis of the features more complicated.

\begin{table}[t]
	\centering
	\begin{tabular}{c c c c c c}
	\hline
	\multicolumn{1}{c}{} & \multicolumn{1}{c}{Proton} & \multicolumn{4}{c}{Wire feature radius (cm)} \\
	Field configuration & energy (MeV) & \multicolumn{1}{c}{60 ns} & \multicolumn{1}{c}{70 ns} & 80 ns & Analytic \\
	\hline
	High-field (13.5 T) & 14.7 &  2.02 $\pm 0.06$ & 2.00 $\pm 0.03$ & 1.95 $\pm 0.03$ & 1.89 \\
	High-field (13.5 T) & 3 &  3.02 $\pm 0.10$ & 2.98 $\pm 0.03$ & 2.86 $\pm 0.10$ & 2.82 \\
	Low-field (9 T) & 14.7 &  1.42 $\pm 0.03$ & 1.42 $\pm 0.03$ & 1.50 $\pm 0.03$ & 1.54 \\
	Low-field (9 T) & 3 &  1.98 $\pm 0.06$ & 2.06 $\pm 0.16$ & 2.18 $\pm 0.06$ & 2.29 \\
	\hline \hline
	\end{tabular}
	\caption{
	The measured radii of the wire feature from the experimental proton images and the analytic expectation using equations (\ref{eq:d14}) and (\ref{eq:d3}) for nominal magnetic field and proton energy parameters.
	The uncertainty in feature radius is determined by the width of the proton intensity maxima for each image.
	}
	\label{Tab:Radii}
\end{table}

We can analytically calculate the extent of the caustic for our system by making a few geometric assumptions.
The curvature of the wire is large relative to the extent of the incoming plasma profile, so we approximate the field around the wire in the experimental plane as that of an infinite, current carrying wire.
Assuming a straight, 380 $\mu$m radius, 4 mm long wire centered 1 cm from the proton source and 16 cm from the image plane, the approximate radial distance of the proton fluence maximum is calculated as
\begin{equation}
	d_{14.7 \text{ MeV}} \approx 1.63\sqrt{B_0/10\text{ T}} \text{ cm},
	\label{eq:d14}
\end{equation}
\begin{equation}
	d_{3 \text{ MeV}} \approx 2.42\sqrt{B_0/10\text{ T}} \text{ cm},
	\label{eq:d3}
\end{equation}
for 14.7 MeV and 3 MeV probe protons, respectively, where $B_0$ is the maximum field amplitude at the wire surface (see appendix for details).
To apply these equations to the distance relative to the object plane (1 cm from the proton source), divide equations (\ref{eq:d14}) and (\ref{eq:d3}) by the 16 times magnification.
Table \ref{Tab:Radii} lists the radii of the wire field features inferred from the experimental proton images, which are consistent with the predictions of equations \ref{eq:d14} and \ref{eq:d3} --- the radii are roughly 50\% larger for 3 MeV images than the 14.7 MeV images, and approximately 25\% larger for 13.5 T maximum fields than for 9 T.
In general, the analytic approximation underpredicts the high-field radii and overpredicts the low-field radii by $<15\%$, with greater error for the low-energy proton images.
This error likely comes from a combination of geometric assumptions about the wire and from assuming a constant velocity across the length of the wire, where discrete time effects would be more important for the lower-energy protons.

\subsection*{Field Reconstruction}

\begin{figure*}[t]
	\centering
	\includegraphics{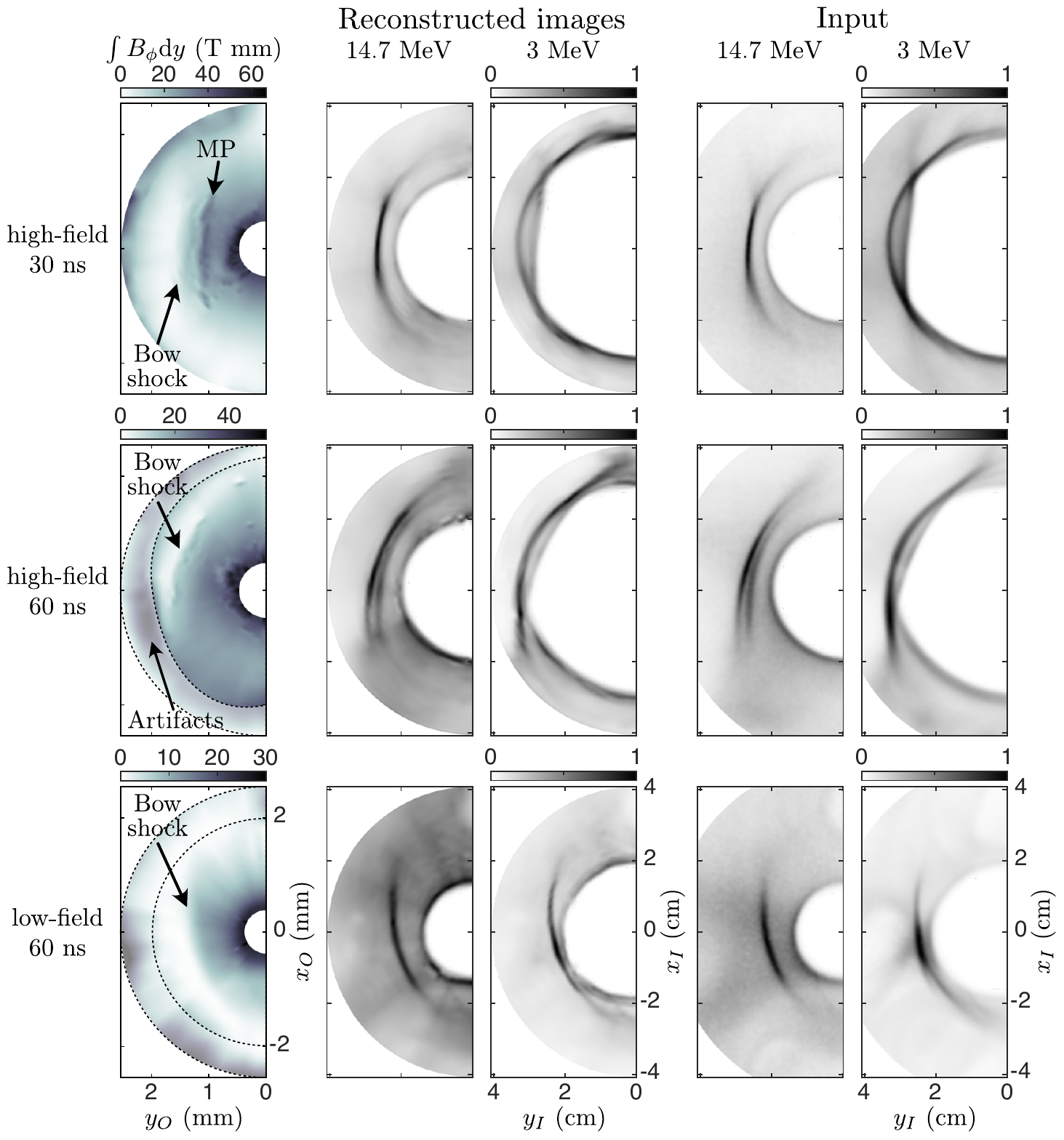}
	\caption{
	Results of the field reconstruction showing the reconstructed path-integrated magnetic field, the synthetic proton images generated from the reconstructed field, and the original images with the same orientation as the images in Figure \ref{Fig:ProtonImages}.
	The field reconstruction uses 40 lineouts of the input image from 0 to $\pi$, assuming that the deflections are all radially outward.
	When stitched together, the reconstructed fields show consistent azimuthal fields that suggest significant compression of the wire field, and we label the features we infer to correspond to the bow shock.
	At 30 ns we also infer a sharp increase in the field that may be a transient magnetopause.
	This reconstruction method assumes a uniform distribution of protons from a point source, and does not handle changes in the proton deflection direction or density scattering, which are likely present in the incoming flow regions.
	Given the limitations of the reconstruction method, the dashed lines outline the region we believe has significant reconstruction errors.
	}
	\label{Fig:ReconstructionImages}
\end{figure*}

\begin{figure}
	\centering
	\includegraphics{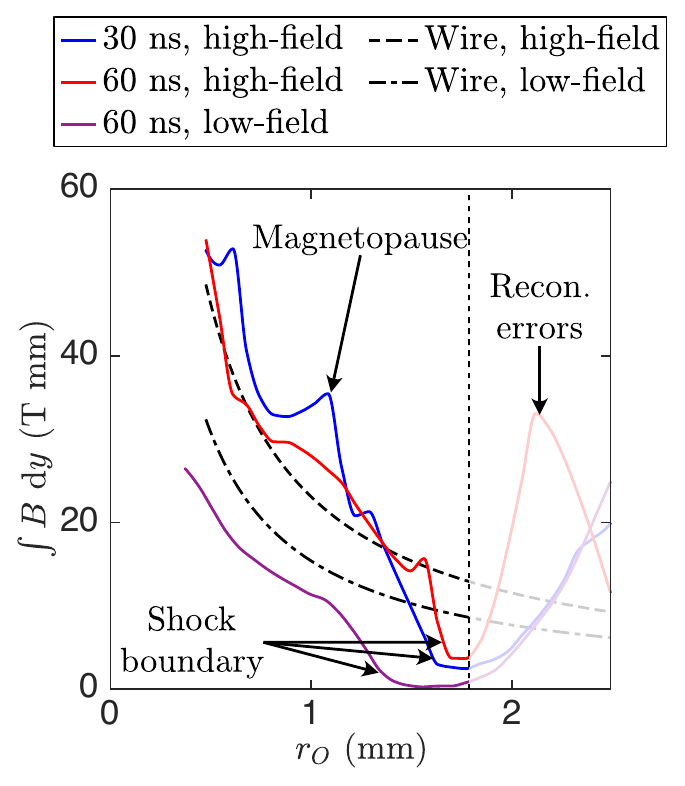}
	\caption{
	Plots of the reconstructed path-integrated fields shown in Figure \ref{Fig:ReconstructionImages}, taken along the centerline of the incoming flow, where the reconstruction should be most accurate.
	All three reconstructed fields show a characteristic decrease of the path-integrated field strength, which occurs around 1.6 mm from the wire for the high field cases, and around 1.3 mm for the low-field case, and we infer to be the shock boundary.
	The reconstructions at 60 ns suggest an increase in the field strength from the nominal wire field, with a pronounced spike at 30 ns that we believe to be a magnetopause.
	The nominal wire field lines assume a 4.5 mm segment of cylindrically symmetric magnetic field from a current-carrying wire.
	The large reconstructed fields at $r>1.9$ mm are an unphysical consequence of the limitations of the reconstruction method.
	}
	\label{Fig:ReconstructedField}
\end{figure}

\begin{figure*}[t]
	\centering
	\includegraphics{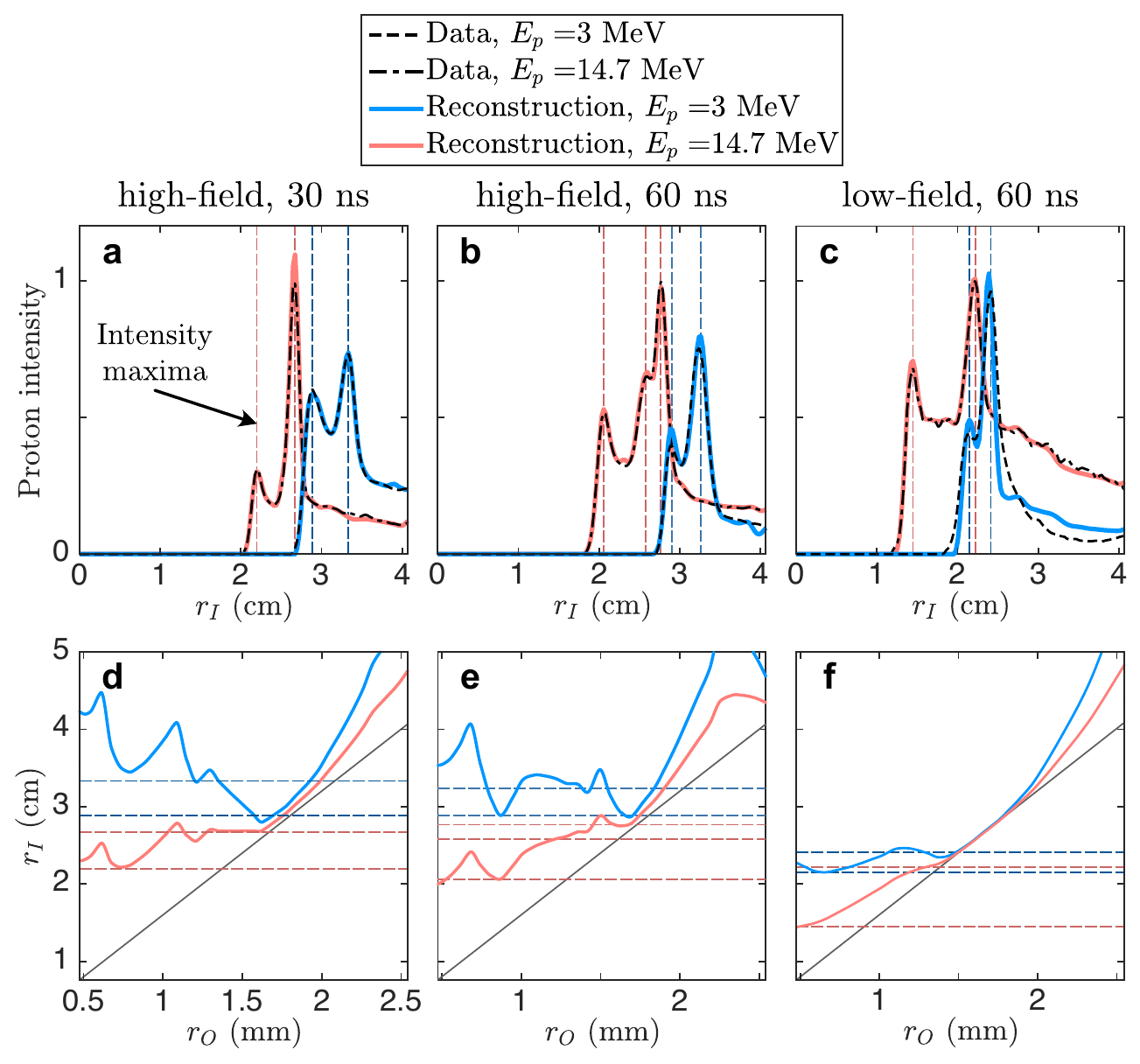}
	\caption{
	(a-c) Lineouts comparing the reconstructed proton intensity to the input intensity from the data shown in Figure \ref{Fig:ReconstructionImages}, along the centerline of the the incoming flow.
	(d-f) Plots of the final image positions of protons as a function of their initial, undeflected positions in the object plane.
	The solid gray lines show the limiting ballistic trajectory.
	The dashed lines correspond to the location of proton intensity peaks in the input data, from which it can be seen how proton deflection features correspond to these image features.
	}
	\label{Fig:ReconstructedImageComparison}
\end{figure*}

The proton images of the wire field are inherently caustic, meaning that the deflections imparted by the magnetic fields cause some regions of neighboring protons to intersect one another before reaching the image plane, and result in regions of greatly increased proton fluence on the image.
There are a number of techniques available for reconstructing magnetic field deflections from proton images\cite{Graziani_RSI2017, Bott_JPP2017, Chen_PRE2017, Kasim_PRE2017, Kasim_PRE2019}, but most are limited to images in the linear contrast regime, and none of those have been shown to deal with images containing strong caustics.
However, because of the inherent symmetry of the magnetic fields in our system primarily deflecting protons radially outward from the wire, we can use the new method of field reconstruction for caustic proton images presented in \citet{Levesque_RSI2021} to estimate the path-integrated magnetic fields of our system.
This reconstruction method takes as input two pseudo-1D proton intensity profiles of the same field system at two distinct proton energies (i.e. 3 MeV and 14.7 MeV) and attempts to create a deflection function from which synthetic proton images fit both input images simultaneously.
The method uses a differential evolution algorithm to iteratively update the perpendicular deflection field of a population of solution candidates to reduce the error between the synthetic images and the input images.
Paraphrasing from \citet{Levesque_RSI2021}, the error of a solution candidate is calculated as
\begin{equation}\label{eq:Error}
	E_{c}(\mathbf{v}_c)^{(t)} = (w_1 \epsilon_1(\mathbf{v}_c)^{(t)}+w_2 \epsilon_2(\mathbf{v}_c)^{(t)})(1 + w_3 \epsilon_3(\mathbf{v}_c)^{(t)}),
\end{equation}
in which $E_c^{(t)}$ is the total heuristic error of candidate $c$ with a vector of deflections $\mathbf{v}_c$ at evolution iteration $t$, and $w_1$, $w_2$, and $w_3$ are scalar weights that act as tuning parameters of the heuristics $\epsilon_1$, $\epsilon_2$.
The individual heuristics are 
\begin{align}
	\epsilon_1(\mathbf{v}_c) &= \text{mean}(|I_{1,\text{recon}}(\mathbf{v}_c)^2-I_1^2|/(I_1+\delta)^2), \label{eq:E1}\\
	\epsilon_2(\mathbf{v}_c) &= \text{mean}(|I_{2,\text{recon}}(\mathbf{v}_c)^2-I_2^2|/(I_2+\delta)^2), \label{eq:E2}\\
	\epsilon_3(\mathbf{v}_c) &= \int_{\alpha_{0,min}}^{\alpha_{0,max}} \left[\frac{\sqrt{1+\left(\partial^2 \Delta\alpha_c(\alpha_0,E_1)/ \partial\alpha_0^2 \right)^2}}{\alpha_{0,max}-\alpha_{0,min}}\right] \partial\alpha_0 -1.  \label{eq:E3}
\end{align}
$\epsilon_1$ and $\epsilon_2$ are the primary heuristics, which essentially determine the difference in the reconstructed images at each energy level $I_{1,recon}$ and $I_{2,recon}$, with the input images $I_1$ and $I_2$; $\delta$ is a small factor to prevent large error where the input proton fluence is zero.
The heuristic $\epsilon_3$ enforces smoothness of the deflection field $\Delta\alpha$ and its gradient, defined based on initial proton trajectories $\alpha_0$.

We note that the reconstruction method of \citet{Levesque_RSI2021} is limited to 1D reconstruction, though our experimental images are not azimuthally symmetric.
Therefore, in order to apply this reconstruction method to our data we must assume that the proton deflections are only in the radial direction.
We justify the assumption of purely radial deflection by noting that the nominal field from the current-carrying wire produces proton deflections almost exclusively in the radial direction, with smaller components due to the system geometry, as illustrated in Figure \ref{Fig:DeflectionIllustration}.
Under the above assumptions, we split the images into 40 radial slices covering $\phi=0$ to $\pi$, setting $r=0$ at the inferred center of the wire in each image.
The slices are then averaged to produce 1D lineouts which are used in the reconstruction, where each slice is solved for independently.

For our reconstruction, the heuristic weights are are mostly taken from the experimental reconstruction presented in \citet{Levesque_RSI2021}, where for the high-field reconstructions $w_1=1$, $w_2=2$, and $w_3=0.5$, and for the low field reconstruction $w_1=0.5$, because the high-field images are typically more informative.
This set of weights prevents the reconstruction from being too stiff to field oscillations and prevents it from getting stuck in higher image-error local minima, which can happen when $w_3$ is large for real images.
We also limit the reconstruction to 31 nodes throughout, and add additional randomness to the candidates when the sum of the relative errors between the reconstruction and the input images, $w_1\epsilon_1+w_2\epsilon_2$, is $\le 0.15$, or when the number of iterations reaches $1.5\times 10^5$, to improve further convergence.

Figure \ref{Fig:ReconstructionImages} shows the results of the field reconstruction for three sets of proton images, displaying the stitched-together path-integrated field corresponding to the reconstructed radial deflections on the left, the reconstructed images in the center, and the input images on the right.
Common to the three reconstructions of Figure \ref{Fig:ReconstructionImages} is a sudden decrease in field strength moving outward from the wire, more easily observed in the lineouts of the path-integrated field displayed in Figure \ref{Fig:ReconstructedField}.
These decreases indicate that the magnetic field is being compressed by the incoming flow and forming a shock surface, with the unmagnetized incoming flow on one side and the magnetosphere of the wire on the other.
The inferred shocks occur at $\approx 1.7$ mm from the wire for the high-field cases, and at $\approx 1.3$ mm for the low-field case, roughly corresponding to the $\approx$3 T surface for the nominal field strengths in both cases, agreeing with inferences of the field from the ITS measurement of a shock at 50 ns.
Given the large standoff distance of the shocks from the wire, and that the lateral extent of these shocks exceeds the size of the wire obstacle, we conclude that these are magnetized bow shocks.
The 30 ns image at high field is the clearest of all images, as the density is expected to be less than $10^{-7}$ g/cm$^3$, and we see the best agreement between the reconstructed images and the input.
Because the plasma at 30 ns is expected to be collisionless, the sharp increase in field 1.1 mm away from the wire, which is consistent across many reconstruction slices, suggests the presence of a magnetopause, if only a transient one.
This magnetopause feature is not observed at later times, possibly because the increasing density and collisionality of the plasma diffuses this feature.

On the topic of density, we chose to only reconstruct the field at 30 and 60 ns because the quality of the images is much better for applying our reconstruction method than those at later times in which we observe increased density scattering, which the reconstruction method does not consider.
Because the reconstruction only considers magnetic field deflections, the additional effects on the proton intensity caused by density scattering and any nonuniformity of the proton beam can only be interpreted by the reconstruction as needing additional radial deflections.
The large increases in field observed toward the edges of the reconstructions which are labeled in Figures \ref{Fig:ReconstructionImages} and \ref{Fig:ReconstructedField} are unphysical, and arise where the reconstruction must move excess protons off of the image to account for those additional factors.

Figure \ref{Fig:ReconstructedImageComparison}(a-c) directly compares the reconstructed proton intensity profiles with the input data along the centerline of the incoming flow, where the compression should be largest, and the assumption of radial deflection should be most applicable.
We see very good agreement between both the 3 and 14.7 MeV images for the two high-field cases, with the best fit at 30 ns.
The reconstruction is not as accurate for the low-field images, possibly because the field cannot hold off the flow as strongly, causing to increased density scattering of the protons closer to the wire.
Corresponding to these lineouts, Figure \ref{Fig:ReconstructedImageComparison}(d-f) shows the proton trajectories --- the final image position $r_I$ as a function of undeflected proton position in the object plane $r_O$ --- for both energy levels.
The flatness and degeneracy of $r_I$ provides a measure of how many protons reach the image around that point.
The dashed lines show the location of peaks in the proton intensity of the input images, and by looking at their intersections with the trajectory plots it is clear that the sharp field decrease (a trough in most of the trajectories) causes the primary shock features in the images. 
Across the shock surface, the protons passing through the magnetized side are deflected further out, and those passing through the region of low field are more ballistic.
These proton trajectories intersect at approximately the same location as the shock, causing the observed increase in proton intensity outside of the wire feature on the 14.7 MeV images, and causing the flat surface inside of the wire feature on the 3 MeV, high-field images, because the shock feature location is effectively inside or coinciding with the wire feature radius.
The double shock feature observed for the high-field case at 60 ns is found to be caused by the slight increase in field at the inside edge of the shock, deflecting some protons farther out than the others that comprise the shock feature.

Using our reconstruction method we have assumed that the proton deflections are only due to interaction with the magnetic field, but could an electric field across a shock interface contribute?
First, note that proton deflections by electric fields are inversely proportional to proton energy, whereas deflections by a magnetic field are inversely proportional to the square root of the proton energy, so magnetic fields will be much more effective at deflecting the high-energy probe protons than electric fields, in general.\cite{Kugland_RSI2012}
Another factor working in our favor is that the overall amplitude of an electric field at this interface should be fairly small, especially when compared to the amplitude of the magnetic field. 
Additionally, the volume of the electric field should be very small since it only exists in the shock layer, so the amplitude of any electric field would have to be much greater than the magnetic field to have a significant influence.
\citet{Hua_PRL2019} reported the presence of fairly strong electric fields ($>300$ V) across a strong shock, at a density of $1.5$ mg/cm$^3$ and a temperature of 140 eV. 
Using the same analytic approximation of the electric field potential as \citet{Hua_PRL2019}, $\Delta \Phi \approx \ln (\rho_2/\rho_1)k_B T_e /e$, for the much lower density and temperature of our system we would expect a potential of $\approx 4$ V across the shock.
Therefore, any electric field that is induced at the shock should only negligibly affect the proton trajectories when compared with the magnetic field.

Aside from what we have been able to reconstruct directly, there are other features on the proton images worth noting.
For example, we can observe the evolution of the incoming flow from the bubble of decreased proton fluence upstream from the wire, caused by increased density scattering.
The increased fluence at the edge of this feature implies the presence of intrinsic electromagnetic fields in the plasma.
This feature expands laterally in time, and at late times the 3 MeV images become partially obscured by the increased density and lateral extent of plasma flowing around the wire.
We can also see that the shock likely extends farther around the wire than indicated by just the leading shock feature, particularly apparent for the images at 70 ns.
Additionally there are some subtler, continuous lines of increased proton fluence in the bottom-right corner of the high-field 3 MeV image at 60 ns.
The angle of these lines changes by approximately $20^{\circ}$ across a surface separating a region of lower proton flux toward the wire.
If these lines correspond to traveling magnetosonic waves, the rotation would indicate magnetic compression across an oblique MHD shock which extends beyond the physical wire obstacle.
Additionally, as the bow shock wraps around the object in time, our assumption of primarily radial proton deflections for our reconstruction technique becomes less applicable.

Our reconstructions are able to find large-scale patterns that cause the shock features on the proton images, but the limitations of reconstructing the field from 1D proton intensity slices with purely radial outward deflections cause the reconstruction to miss many of the smaller-scale features.
In reality, we expect that there are additional components of the field that drive the smaller features, like the decreased proton fluence within the bubble of the incoming flow, and also the smaller bubbles near the edges of the images as in the low-field, 60 ns images.
Indeed, if the field was causing purely radial outward deflections, then the field would be expected to have azimuthal symmetry throughout, which is not the case in our reconstructions.
However, adapting the DE reconstruction to two dimensions incurs significant additional complexity and much-increased computational cost, and will require substantial effort to mature. 
Therefore, the reconstructed fields we present should be considered as semi-quantitative, providing estimates of the compression of the field under the assumption that at any slice the deflection is primarily radial, due to the nominal strength of the field around the wire.

\section{MHD Simulations} \label{Sec:MHDSimulations}

\begin{figure}
	\centering
	\includegraphics{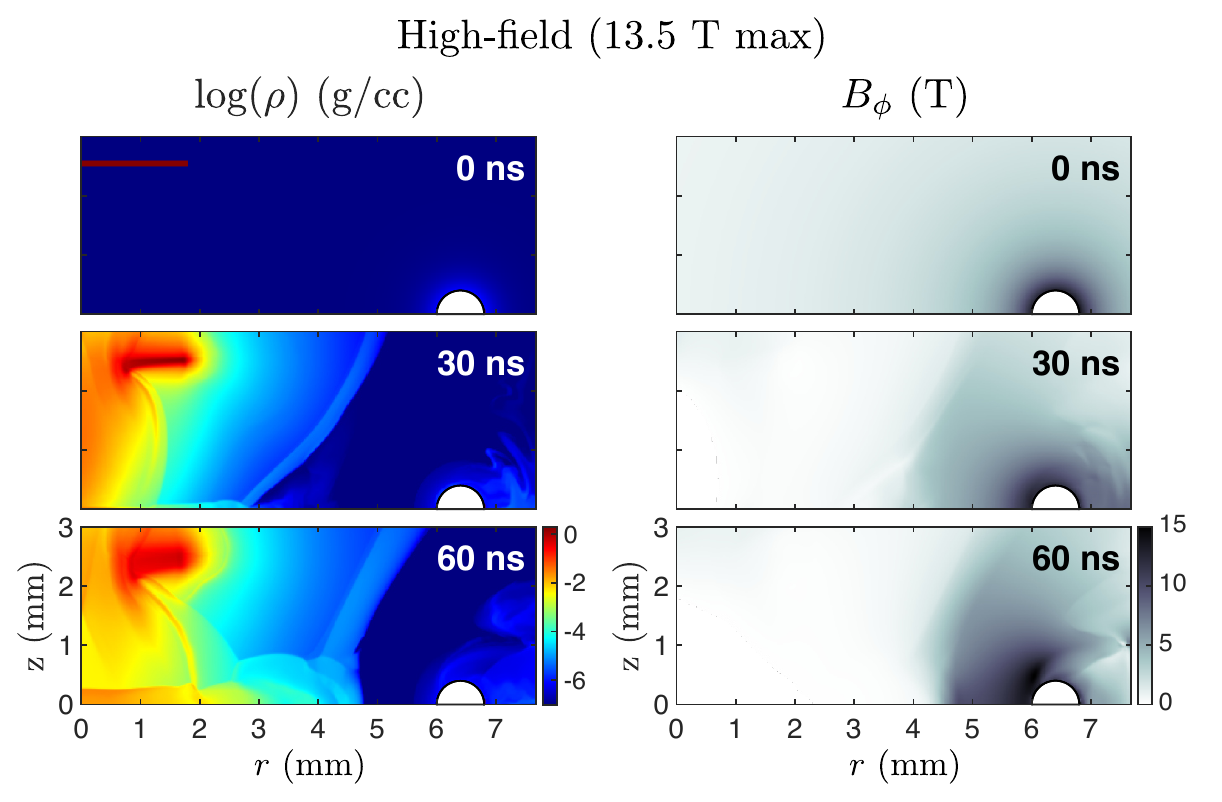}
	\caption{
	Time series of an MHD simulation of the high-field system with a maximum field of 13.5 T, allowing for rear surface blowoff.
	Left: density, where the plasma propagates axially, collides with the $z=0$ boundary, and is redirected as expected.
	Right: magnetic field in the $\phi$ direction, showing significant compression of the magnetic field upstream of the wire and redirecting the field as the flow moves downstream around it.
	The outline of the magnetic field boundary can be seen at 30 ns, and is still present at 60 ns.
	}
	\label{Fig:HighFieldSim}
\end{figure}

\begin{figure}
	\centering
	\includegraphics{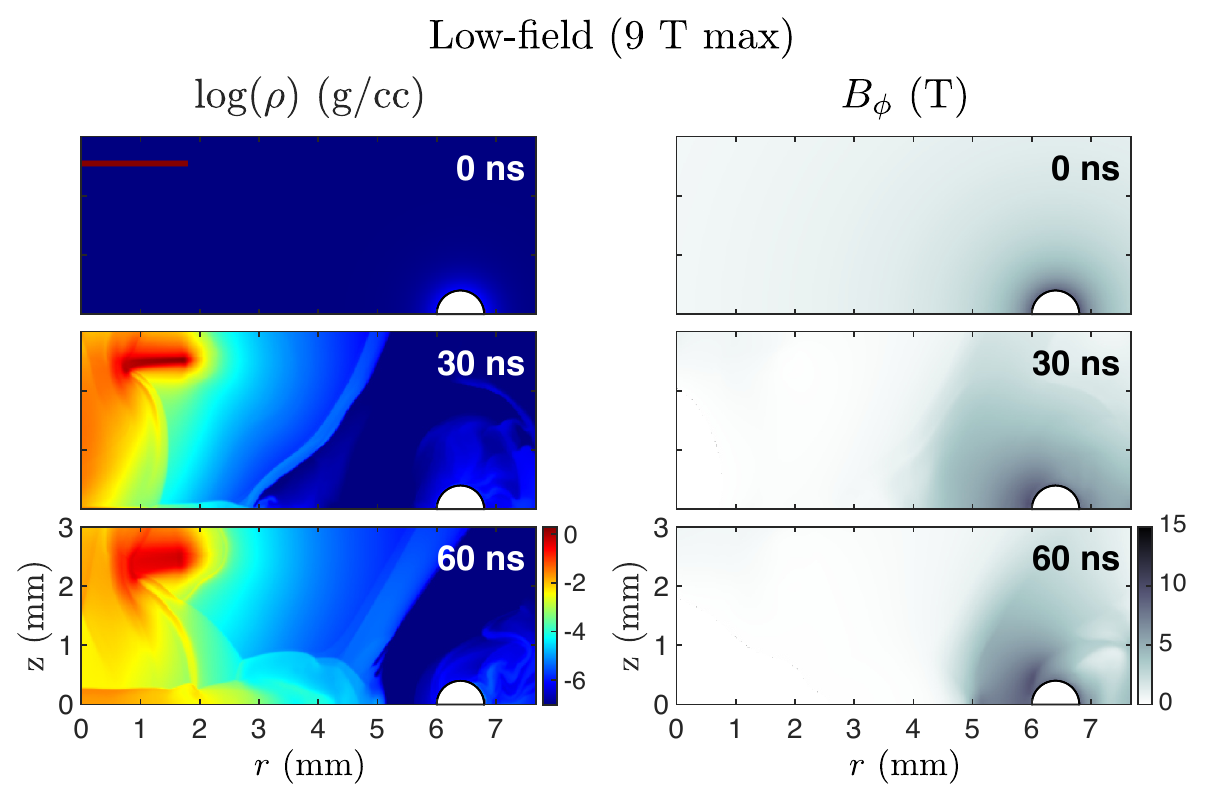}
	\caption{
	Time series of an MHD simulation of the low-field system with a maximum field of 9 T, allowing for rear surface blowoff.
	The results are at 30 ns are similar to what is seen in the high-field simulation.
	The flow is able to push farther in at 60 ns with the lower field, and the shock is approximately 0.4 mm closer to the wire than the high-field case.
	}
	\label{Fig:LowFieldSim}
\end{figure}

We also performed 2D MHD simulations of the system at both nominal field amplitudes using the FLASH code.
The simulation system is defined in a 2D cylindrical geometry with the $z$ axis defined at the center of the carbon targets as in the colliding target simulations of Section \ref{Sec:ExperimentalSetup}.
Figures \ref{Fig:HighFieldSim} and \ref{Fig:LowFieldSim} show the density and magnetic field results of these simulations at 0, 30, and 60 ns.
After the initial 1 ns laser drive, the target plasma expands along the axis of laser incidence as expected, and also laterally, generating a shock in the ambient medium which quickly propagates outward.
By 30 ns the main plasma plume has already reached the collision plane, and is being redirected radially outward in the same way as in Figure \ref{Fig:SourceComp}.
This outflow sweeps up the magnetic field and acts against an increasing magnetic pressure as it travels, forming a shock and a region closer to the wire in which magnetic pressure dominates and redirects the incoming flow.

At 60 ns, we see that the boundary between the incoming flow and the inner, magnetized region occurs $\approx 1.9$ mm and $\approx 1.4$ mm from the center of the wire for the high-field and low-field cases, and the peak density occurs $\approx 200$ $\mu$m upstream.
These boundaries are farther from the wire than the features we see in the reconstructed fields of Figure \ref{Fig:ReconstructionImages}, although it is important to note that we are comparing a slice of the field with the path-integrated values.
Inside of the magnetized region close to the wire the plasma is effectively evacuated, dropping to the minimum simulation density.
The density of the region immediately upstream of the separation reaches $\approx 10^{-5}$ g/cc corresponding to $n_e \approx 3\times 10^{18}$ cm$^{-3}$, which is notably lower than the upstream density inferred from the 50 ns ITS measurement of Figure \ref{Fig:ITS}.
These issues with density may be a result of the cylindrical geometry not allowing the flow to properly pass the wire, and resulting in a greater influence of the magnetic field.

Note that we do not observe agreement between the magnetic fields from FLASH and the reconstructed field at 30 ns.
In the simulations, the shock is instead first set up by a shock in the background material originating from the laser spot which quickly expands outward toward the wire.
This shock is likely the result of the high ambient density in the simulation relative to the experiment, and may not be physical.
After this shock reaches the wire it is reflected back toward the axis of flow, and acts as the seed for the sustained discontinuity at later times $\geq 30$ ns.
When a boundary is imposed on the back side of the target to block this shock in the ambient medium as was done for the hydrodynamic simulations in Figure \ref{Fig:SourceComp}, the colliding flows remain much more collimated, and do not seem to set up a shock as the flow approaches the wire, instead plowing through the increased field despite reaching approximately the same conditions as when there is no boundary.
In reality, at early times the shock may be initially set up by the faster, more diffuse, collisionless plasma streaming ahead of the primary flow, which cannot be captured by a hydrodynamic code.
A hybrid fluid-kinetic code may be needed to more accurately model this system throughout its evolution.
For any future experiments, efforts should be made to keep the system collisionless for longer, enabling more direct simulations using fully kinetic codes.

The imposed cylindrical geometry of these simulations likely causes some of the observed discrepancies between the simulations and the data.
In the real system the wire only has a limited extent, and is curved away from the carbon targets, rather than around them as the cylindrical simulation geometry implies.
However, for 2D simulations the cylindrical geometry is necessary to properly capture the dynamics of the expanding plasma flows --- a 2D Cartesian geometry results in much greater plasma density.
Additionally, the magnetic field from the wire is initialized corresponding to the 2D Cartesian field around an infinite current-carrying wire.
The mismatch of field geometry may cause a stronger effective field to act on the plasma, and may reduce the compression.
The cylindrical geometry is likely the optimal 2D implementation of this system, but we expect that any 2D simulation cannot properly capture the dynamics of the plasma flowing around the obstacle in the real system.
Although moving to 3D simulations may alleviate some issues, they are very resource-intensive because of the large system volume.
Another caveat of translating the experiment to MHD simulations is that FLASH does not include any means of driving or calculating the current flowing through the wire, so to make up for diffusion of the field out of the wire, the field inside the wire is replenished to the initial state at each time step, and this inhibits field pileup on or in the wire as the flow compresses the field.

\section{Discussion}\label{Sec:Discussion}

\begin{table}[t]
\begin{ruledtabular}
	\centering
	\begin{tabular}{l c c}
		Parameter & Solar Wind & Experiment \\
		\colrule
		$L$ (mm) & $6.4 \times 10^{10}$ & $\sim 1$ \\
		$\rho$ (g/cm$^3$) & $\sim 10^{-23}$ & $10^{-6}-<10^{-4}$ \\
		$\bar{v}$ (km/s) & 400 & 150 \\
		$B$ (T) & $\sim 10^{-9}$ & $\leq 13.5$ \\
		$T$ (eV) & 12 & 6 \\
		$v_A$ (km/s) & $\sim 10$ & 80 \\
		$c_s$ (km/s) & 45 & $15.5$ \\
		$M_{fms}$ & $\sim 9$ & $\sim 2$ \\
		$R_m$ & $\gg 1$ & $\sim 4$ \\
	\end{tabular}
	\end{ruledtabular}
	\caption{
	Comparison of average parameters in the interaction of the solar wind with the Earth's magnetosphere and experimental parameters, where $L$ is the characteristic system length scale, $\rho$ is the plasma density, $\bar{v}$ is the mean velocity, $B$ is the external magnetic field, $T$ is the temperature, $v_A$ is the Alfv\'en speed, $c_s$ is the sound speed, $M_{fms}$ is the fast magnetosonic Mach number, and $R_m$ is the magnetic Reynolds number.
	}
	\label{Tab:ParameterComparison}
\end{table}

We have observed signatures of a shock at a significant standoff distance from our magnetized wire in both our high-field Thomson scattering measurement at 50 ns, and multiple proton images at both high-field and low-field between 30 and 80 ns.
Using these measurements, we can now further inspect some of the properties of our system and compare them to the Earth's magnetosphere.
Table \ref{Tab:ParameterComparison} provides some average parameters of the solar wind interaction with the Earth's magnetosphere\cite{Hundhausen_Chapter,Mullan_SP2006} and from our experiment.
Looking first at the $\beta_{ram}$, it is apparent that the orders of magnitude difference between the ram pressure are balanced by the corresponding change in magnitude of the field strength as designed.
The experimental Alfv\'en speed exceeds the sound speed unlike in the solar wind, but the flow still exceeds the fast magnetosonic speed, which indicates that a shock should form.

We now address the other constraints listed Section \ref{Sec:Introduction}. 
The colliding plasma flow source was able to exist in a shock-favorable regime for at least 20 ns in the high-field case, allowing for a few wire-crossings in that time, satisfying (i).
We observe the resulting transient magnetopause in the proton images.
Turning to magnetic diffusion, using the measured $\sim 6$ eV temperature and a velocity of 150 km/s predicted by the FLASH simulations over a characteristic 1 mm, we estimate a magnetic Reynolds number $R_m \approx 4$, which satisfies constraint (ii). 
Although $R_m>1$ and advection dominates over longer length and time scales, diffusion is expected to dominate at length scales $<250$ $\mu$m, which agrees roughly with the thickness of the sheath region in our MHD simulations.
The system is highly dynamic, however, so the magnetic Reynolds number may not fully capture the effects of changing gradients in the field.
We clearly see in the proton images that the field is advected as the plasma flows around the wire, but diffusion may set the lower bound on shock thickness and act to reduce the maximum compression of the field.

As mentioned previously, from around 50 ns onward, the plasma interaction in the experiment is expected to be highly collisional, so the primary scale of shock formation is the collisional mean free path.
At the predicted $5 \times 10^{-6}$ g/cm$^3$ density, $150$ km/s velocity, and 6 eV temperature, the electron mean free path is only $\sim$25 nm and the ion mean free path $\sim$200 nm, much smaller than either the gyroradius or the obstacle scale, which satisfies the final constraint (iii).
However, the measured density from ITS is approximately an order of magnitude greater than the hydrodynamic simulations.
The most likely explanation for this discrepancy is that the additional magnetic field pileup slows the flow and causes the density to increase just ahead of the shock, but even at this density the resulting mean free paths are still $<1$ $\mu$m, so shock formation is still expected.
We have thus met the requirements for magnetized shock formation in this system.
However, these results in the collisional regime cannot be directly related to a collisionless system like the Earth's bow shock.
In a collisionless system the magnetopause should remain very distinct and display significant compression of the field, like we infer exists in our experiment at 30 ns.

\section{Conclusions}\label{Sec:Conclusion}
We performed a series of experiments at the OMEGA laser facility which successfully generated standing bow shocks from the interaction of a flowing plasma with a strong magnetic pressure.
The expanding plasma flow created by the collision of two laser-irradiated plasma plumes attained the plasma parameters necessary to produce a bow shock in the interaction with the magnetic field surrounding a current-carrying wire.
We infer the presence of an MHD shock at a significant standoff distance from the magnetized wire, measured with both Thomson scattering spectrometry and proton imaging diagnostics.
Using the spatially resolved Imaging Thomson Scattering diagnostic we measured a sharp doubling of the electron number density in a shock at a standoff distance of roughly 1.8 mm from the wire, 50 ns after initializing the plasma flow.
This same diagnostic also measured optical emission spectra of the plasma, unperturbed by the Thomson scattering probe laser, which are fit well by a 6 eV carbon plasma, in agreement with the temperature predicted in the FLASH simulations.
Based on the density jump in the shock, the plasma temperature, and the nominal magnetic field strength, we determine that this shock must be a significantly magnetized, fast magnetosonic shock.

Proton images of the system show features corresponding to the shadow of the wire structure, the nominal magnetic field profile, and the compressed magnetic field of the shock.
Owing to the quasi-two dimensional system geometry, we analytically estimated the size of the proton image feature caused by the nominal magnetic field around the wire and compared these estimates to the data.
Sharp proton intensity features upstream of the wire field feature correspond to compression of the magnetic field by the inflowing plasma, which we confirmed by reconstructions of the path-integrated magnetic fields, and indicate the formation of a bow shock.
Based on proton images, the shocks appear to be short lived, likely due to the rapidly increasing density of the incoming plasma.
The clarity of the large scale features on the proton images is notable, and demonstrates that this diagnostic is an excellent way to image shocks in a magnetized plasma, and there are some interesting features that would benefit from further investigation.
To improve the quality of magnetic field reconstructions from proton images in future experiments, steps should be taken to reduce the curvature of the applied field and make the system as two-dimensional as possible.
With further tuning of the experimental setup, by reducing the density or velocity of the plasma flow to remain in the collisionless regime for longer, and possibly increasing the temperature to decrease the magnetic diffusivity, these experiments could be a significant tool for studying physics relevant to the formation and evolution of planetary bow shocks.

\section{Acknowledgements}

This work was funded by the DOE, through the NNSA Center of Excellence under grant number DE-NA0003869, the NNSA-DP and SC-OFES Joint Program in HEDLP, grant number DE-NA0002956, the NLUF Program and Rice University, grant number DE-NA0002722, NLUF Program, grant number DE-NA0002719, and through the LLE, University of Rochester by the NNSA/OICF under Cooperative Agreement numbers DE-NA0001944 and DE-NA0003856.
Partial support for this work was also provided by NASA through Einstein Postdoctoral Fellowship Grant No. PF3-140111 awarded by the Chandra X-ray Center, which is operated by the Astrophysical Observatory for NASA under Contract No. NAS8-03060. 
This work was also partially supported by the U.S. Department of Energy under Field Work Proposal No. 57789 to Argonne National Laboratory, Subcontract No. 536203 with Los Alamos National Laboratory, and Subcontract B632670 with Lawrence Livermore National Laboratory to the University of Chicago. 
The FLASH software used in this work was developed in part by the DOE NNSA ASC- and DOE Office of Science ASCR-supported Flash Center for Computational Science at the University of Chicago.
This work was also supported by the U.S. Department of Energy through the Los Alamos National Laboratory. Los Alamos National Laboratory is operated by Triad National Security, LLC, for the National Nuclear Security Administration of U.S. Department of Energy (Contract No. 89233218CNA000001).

\section*{Supplementary Material}
The additional ITS measurements referred to in Section \ref{Sec:ThomsonScattering} are included in the Supplementary Material file.

\section*{Data Availability}
The data that support the findings of this article are available from the corresponding author upon reasonable request.

\bibliography{References}

\begin{thebibliography}{45}%
\makeatletter
\providecommand \@ifxundefined [1]{%
 \@ifx{#1\undefined}
}%
\providecommand \@ifnum [1]{%
 \ifnum #1\expandafter \@firstoftwo
 \else \expandafter \@secondoftwo
 \fi
}%
\providecommand \@ifx [1]{%
 \ifx #1\expandafter \@firstoftwo
 \else \expandafter \@secondoftwo
 \fi
}%
\providecommand \natexlab [1]{#1}%
\providecommand \enquote  [1]{``#1''}%
\providecommand \bibnamefont  [1]{#1}%
\providecommand \bibfnamefont [1]{#1}%
\providecommand \citenamefont [1]{#1}%
\providecommand \href@noop [0]{\@secondoftwo}%
\providecommand \href [0]{\begingroup \@sanitize@url \@href}%
\providecommand \@href[1]{\@@startlink{#1}\@@href}%
\providecommand \@@href[1]{\endgroup#1\@@endlink}%
\providecommand \@sanitize@url [0]{\catcode `\\12\catcode `\$12\catcode
  `\&12\catcode `\#12\catcode `\^12\catcode `\_12\catcode `\%12\relax}%
\providecommand \@@startlink[1]{}%
\providecommand \@@endlink[0]{}%
\providecommand \url  [0]{\begingroup\@sanitize@url \@url }%
\providecommand \@url [1]{\endgroup\@href {#1}{\urlprefix }}%
\providecommand \urlprefix  [0]{URL }%
\providecommand \Eprint [0]{\href }%
\providecommand \doibase [0]{https://doi.org/}%
\providecommand \selectlanguage [0]{\@gobble}%
\providecommand \bibinfo  [0]{\@secondoftwo}%
\providecommand \bibfield  [0]{\@secondoftwo}%
\providecommand \translation [1]{[#1]}%
\providecommand \BibitemOpen [0]{}%
\providecommand \bibitemStop [0]{}%
\providecommand \bibitemNoStop [0]{.\EOS\space}%
\providecommand \EOS [0]{\spacefactor3000\relax}%
\providecommand \BibitemShut  [1]{\csname bibitem#1\endcsname}%
\let\auto@bib@innerbib\@empty
\bibitem [{\citenamefont {Spreiter}, \citenamefont {Summers},\ and\
  \citenamefont {Alksne}(1966)}]{Spreiter_PSS1966}%
  \BibitemOpen
  \bibfield  {author} {\bibinfo {author} {\bibfnamefont {J.~R.}\ \bibnamefont
  {Spreiter}}, \bibinfo {author} {\bibfnamefont {A.~L.}\ \bibnamefont
  {Summers}},\ and\ \bibinfo {author} {\bibfnamefont {A.~Y.}\ \bibnamefont
  {Alksne}},\ }\bibfield  {title} {\enquote {\bibinfo {title} {Hydromagnetic
  flow around the magnetosphere},}\ }\href
  {https://doi.org/10.1016/0032-0633(66)90124-3} {\bibfield  {journal}
  {\bibinfo  {journal} {Planetary and Space Science}\ }\textbf {\bibinfo
  {volume} {14}},\ \bibinfo {pages} {223--253} (\bibinfo {year}
  {1966})}\BibitemShut {NoStop}%
\bibitem [{\citenamefont {Fairfield}(1971)}]{Fairfield_JGR1971}%
  \BibitemOpen
  \bibfield  {author} {\bibinfo {author} {\bibfnamefont {D.~H.}\ \bibnamefont
  {Fairfield}},\ }\bibfield  {title} {\enquote {\bibinfo {title} {Average and
  unusual locations of the earth's magnetopause and bow shock},}\ }\href
  {https://doi.org/10.1029/JA076i028p06700} {\bibfield  {journal} {\bibinfo
  {journal} {Journal of Geophysical Research (1896-1977)}\ }\textbf {\bibinfo
  {volume} {76}},\ \bibinfo {pages} {6700--6716} (\bibinfo {year}
  {1971})}\BibitemShut {NoStop}%
\bibitem [{\citenamefont {Zhuang}\ and\ \citenamefont
  {Russell}(1981)}]{Zhuang_JGR1981}%
  \BibitemOpen
  \bibfield  {author} {\bibinfo {author} {\bibfnamefont {H.~C.}\ \bibnamefont
  {Zhuang}}\ and\ \bibinfo {author} {\bibfnamefont {C.~T.}\ \bibnamefont
  {Russell}},\ }\bibfield  {title} {\enquote {\bibinfo {title} {An analytic
  treatment of the structure of the bow shock and magnetosheath},}\ }\href
  {https://doi.org/10.1029/JA086iA04p02191} {\bibfield  {journal} {\bibinfo
  {journal} {Journal of Geophysical Research: Space Physics}\ }\textbf
  {\bibinfo {volume} {86}},\ \bibinfo {pages} {2191--2205} (\bibinfo {year}
  {1981})}\BibitemShut {NoStop}%
\bibitem [{\citenamefont {Cairns}\ and\ \citenamefont
  {Grabbe}(1994)}]{CairnsGRL1994}%
  \BibitemOpen
  \bibfield  {author} {\bibinfo {author} {\bibfnamefont {I.~H.}\ \bibnamefont
  {Cairns}}\ and\ \bibinfo {author} {\bibfnamefont {C.~L.}\ \bibnamefont
  {Grabbe}},\ }\bibfield  {title} {\enquote {\bibinfo {title} {Towards an mhd
  theory for the standoff distance of earth's bow shock},}\ }\href
  {https://doi.org/10.1029/94GL02551} {\bibfield  {journal} {\bibinfo
  {journal} {Geophysical Research Letters}\ }\textbf {\bibinfo {volume} {21}},\
  \bibinfo {pages} {2781--2784} (\bibinfo {year} {1994})}\BibitemShut {NoStop}%
\bibitem [{\citenamefont {Luhmann}(1995)}]{Luhmann_Chapter}%
  \BibitemOpen
  \bibfield  {author} {\bibinfo {author} {\bibfnamefont {J.~G.}\ \bibnamefont
  {Luhmann}},\ }\bibfield  {title} {\enquote {\bibinfo {title} {Plasma
  interactions with unmagnetized bodies},}\ }in\ \href@noop {} {\emph {\bibinfo
  {booktitle} {Introduction to Space Physics}}},\ \bibinfo {editor} {edited by\
  \bibinfo {editor} {\bibnamefont {{M. G. Kivelson and C. T. Russell}}}}\
  (\bibinfo  {publisher} {Cambridge University Press},\ \bibinfo {year}
  {1995})\ Chap.~\bibinfo {chapter} {8}, pp.\ \bibinfo {pages}
  {203--226}\BibitemShut {NoStop}%
\bibitem [{\citenamefont {Khodachenko}\ \emph {et~al.}(2007)\citenamefont
  {Khodachenko}, \citenamefont {Ribas}, \citenamefont {Lammer}, \citenamefont
  {Grie{\ss}meier}, \citenamefont {Leitner}, \citenamefont {Selsis},
  \citenamefont {Eiroa}, \citenamefont {Hanslmeier}, \citenamefont {Biernat},
  \citenamefont {Farrugia},\ and\ \citenamefont
  {Rucker}}]{Khodachenko_Astrobiology2007}%
  \BibitemOpen
  \bibfield  {author} {\bibinfo {author} {\bibfnamefont {M.~L.}\ \bibnamefont
  {Khodachenko}}, \bibinfo {author} {\bibfnamefont {I.}~\bibnamefont {Ribas}},
  \bibinfo {author} {\bibfnamefont {H.}~\bibnamefont {Lammer}}, \bibinfo
  {author} {\bibfnamefont {J.-M.}\ \bibnamefont {Grie{\ss}meier}}, \bibinfo
  {author} {\bibfnamefont {M.}~\bibnamefont {Leitner}}, \bibinfo {author}
  {\bibfnamefont {F.}~\bibnamefont {Selsis}}, \bibinfo {author} {\bibfnamefont
  {C.}~\bibnamefont {Eiroa}}, \bibinfo {author} {\bibfnamefont
  {A.}~\bibnamefont {Hanslmeier}}, \bibinfo {author} {\bibfnamefont {H.~K.}\
  \bibnamefont {Biernat}}, \bibinfo {author} {\bibfnamefont {C.~J.}\
  \bibnamefont {Farrugia}},\ and\ \bibinfo {author} {\bibfnamefont {H.~O.}\
  \bibnamefont {Rucker}},\ }\bibfield  {title} {\enquote {\bibinfo {title}
  {Coronal mass ejection (cme) activity of low mass m stars as an important
  factor for the habitability of terrestrial exoplanets. i. cme impact on
  expected magnetospheres of earth-like exoplanets in close-in habitable
  zones},}\ }\href {https://doi.org/10.1089/ast.2006.0127} {\bibfield
  {journal} {\bibinfo  {journal} {Astrobiology}\ }\textbf {\bibinfo {volume}
  {7}},\ \bibinfo {pages} {167--184} (\bibinfo {year} {2007})}\BibitemShut
  {NoStop}%
\bibitem [{\citenamefont {{Vidotto}}\ \emph {et~al.}(2015)\citenamefont
  {{Vidotto}}, \citenamefont {{Fares}}, \citenamefont {{Jardine}},
  \citenamefont {{Moutou}},\ and\ \citenamefont
  {{Donati}}}]{Vidotto_MNRAS2015}%
  \BibitemOpen
  \bibfield  {author} {\bibinfo {author} {\bibfnamefont {A.~A.}\ \bibnamefont
  {{Vidotto}}}, \bibinfo {author} {\bibfnamefont {R.}~\bibnamefont {{Fares}}},
  \bibinfo {author} {\bibfnamefont {M.}~\bibnamefont {{Jardine}}}, \bibinfo
  {author} {\bibfnamefont {C.}~\bibnamefont {{Moutou}}},\ and\ \bibinfo
  {author} {\bibfnamefont {J.~F.}\ \bibnamefont {{Donati}}},\ }\bibfield
  {title} {\enquote {\bibinfo {title} {{On the environment surrounding close-in
  exoplanets}},}\ }\href {https://doi.org/10.1093/mnras/stv618} {\bibfield
  {journal} {\bibinfo  {journal} {MNRAS}\ }\textbf {\bibinfo {volume} {449}},\
  \bibinfo {pages} {4117--4130} (\bibinfo {year} {2015})}\BibitemShut {NoStop}%
\bibitem [{\citenamefont {T{\'o}th}\ \emph {et~al.}(2005)\citenamefont
  {T{\'o}th}, \citenamefont {Sokolov}, \citenamefont {Gombosi}, \citenamefont
  {Chesney}, \citenamefont {Clauer}, \citenamefont {De~Zeeuw}, \citenamefont
  {Hansen}, \citenamefont {Kane}, \citenamefont {Manchester}, \citenamefont
  {Oehmke}, \citenamefont {Powell}, \citenamefont {Ridley}, \citenamefont
  {Roussev}, \citenamefont {Stout}, \citenamefont {Volberg}, \citenamefont
  {Wolf}, \citenamefont {Sazykin}, \citenamefont {Chan}, \citenamefont {Yu},\
  and\ \citenamefont {K{\'o}ta}}]{Toth_JGR2005}%
  \BibitemOpen
  \bibfield  {author} {\bibinfo {author} {\bibfnamefont {G.}~\bibnamefont
  {T{\'o}th}}, \bibinfo {author} {\bibfnamefont {I.~V.}\ \bibnamefont
  {Sokolov}}, \bibinfo {author} {\bibfnamefont {T.~I.}\ \bibnamefont
  {Gombosi}}, \bibinfo {author} {\bibfnamefont {D.~R.}\ \bibnamefont
  {Chesney}}, \bibinfo {author} {\bibfnamefont {C.~R.}\ \bibnamefont {Clauer}},
  \bibinfo {author} {\bibfnamefont {D.~L.}\ \bibnamefont {De~Zeeuw}}, \bibinfo
  {author} {\bibfnamefont {K.~C.}\ \bibnamefont {Hansen}}, \bibinfo {author}
  {\bibfnamefont {K.~J.}\ \bibnamefont {Kane}}, \bibinfo {author}
  {\bibfnamefont {W.~B.}\ \bibnamefont {Manchester}}, \bibinfo {author}
  {\bibfnamefont {R.~C.}\ \bibnamefont {Oehmke}}, \bibinfo {author}
  {\bibfnamefont {K.~G.}\ \bibnamefont {Powell}}, \bibinfo {author}
  {\bibfnamefont {A.~J.}\ \bibnamefont {Ridley}}, \bibinfo {author}
  {\bibfnamefont {I.~I.}\ \bibnamefont {Roussev}}, \bibinfo {author}
  {\bibfnamefont {Q.~F.}\ \bibnamefont {Stout}}, \bibinfo {author}
  {\bibfnamefont {O.}~\bibnamefont {Volberg}}, \bibinfo {author} {\bibfnamefont
  {R.~A.}\ \bibnamefont {Wolf}}, \bibinfo {author} {\bibfnamefont
  {S.}~\bibnamefont {Sazykin}}, \bibinfo {author} {\bibfnamefont
  {A.}~\bibnamefont {Chan}}, \bibinfo {author} {\bibfnamefont {B.}~\bibnamefont
  {Yu}},\ and\ \bibinfo {author} {\bibfnamefont {J.}~\bibnamefont {K{\'o}ta}},\
  }\bibfield  {title} {\enquote {\bibinfo {title} {Space weather modeling
  framework: A new tool for the space science community},}\ }\href
  {https://doi.org/10.1029/2005JA011126} {\bibfield  {journal} {\bibinfo
  {journal} {Journal of Geophysical Research: Space Physics}\ }\textbf
  {\bibinfo {volume} {110}} (\bibinfo {year} {2005}),\
  10.1029/2005JA011126}\BibitemShut {NoStop}%
\bibitem [{\citenamefont {Ryutov}, \citenamefont {Drake},\ and\ \citenamefont
  {Remington}(2000)}]{Ryutov_ApJS2000}%
  \BibitemOpen
  \bibfield  {author} {\bibinfo {author} {\bibfnamefont {D.~D.}\ \bibnamefont
  {Ryutov}}, \bibinfo {author} {\bibfnamefont {R.~P.}\ \bibnamefont {Drake}},\
  and\ \bibinfo {author} {\bibfnamefont {B.~A.}\ \bibnamefont {Remington}},\
  }\bibfield  {title} {\enquote {\bibinfo {title} {Criteria for scaled
  laboratory simulations of astrophysical {MHD} phenomena},}\ }\href
  {https://doi.org/10.1086/313320} {\bibfield  {journal} {\bibinfo  {journal}
  {The Astrophysical Journal Supplement Series}\ }\textbf {\bibinfo {volume}
  {127}},\ \bibinfo {pages} {465--468} (\bibinfo {year} {2000})}\BibitemShut
  {NoStop}%
\bibitem [{\citenamefont {Ryutov}\ \emph {et~al.}(2001)\citenamefont {Ryutov},
  \citenamefont {Remington}, \citenamefont {Robey},\ and\ \citenamefont
  {Drake}}]{Ryutov_PoP2001}%
  \BibitemOpen
  \bibfield  {author} {\bibinfo {author} {\bibfnamefont {D.~D.}\ \bibnamefont
  {Ryutov}}, \bibinfo {author} {\bibfnamefont {B.~A.}\ \bibnamefont
  {Remington}}, \bibinfo {author} {\bibfnamefont {H.~F.}\ \bibnamefont
  {Robey}},\ and\ \bibinfo {author} {\bibfnamefont {R.~P.}\ \bibnamefont
  {Drake}},\ }\bibfield  {title} {\enquote {\bibinfo {title}
  {Magnetohydrodynamic scaling: From astrophysics to the laboratory},}\ }\href
  {https://doi.org/10.1063/1.1344562} {\bibfield  {journal} {\bibinfo
  {journal} {Physics of Plasmas}\ }\textbf {\bibinfo {volume} {8}},\ \bibinfo
  {pages} {1804--1816} (\bibinfo {year} {2001})}\BibitemShut {NoStop}%
\bibitem [{\citenamefont {Ryutov}\ and\ \citenamefont
  {Remington}(2002)}]{Ryutov_PPCF2002}%
  \BibitemOpen
  \bibfield  {author} {\bibinfo {author} {\bibfnamefont {D.~D.}\ \bibnamefont
  {Ryutov}}\ and\ \bibinfo {author} {\bibfnamefont {B.~A.}\ \bibnamefont
  {Remington}},\ }\bibfield  {title} {\enquote {\bibinfo {title} {Scaling
  astrophysical phenomena to high-energy-density laboratory experiments},}\
  }\href {https://doi.org/10.1088/0741-3335/44/12b/328} {\bibfield  {journal}
  {\bibinfo  {journal} {Plasma Physics and Controlled Fusion}\ }\textbf
  {\bibinfo {volume} {44}},\ \bibinfo {pages} {B407--B423} (\bibinfo {year}
  {2002})}\BibitemShut {NoStop}%
\bibitem [{\citenamefont {Lavraud}\ and\ \citenamefont
  {Borovsky}(2008)}]{Lavraud2008}%
  \BibitemOpen
  \bibfield  {author} {\bibinfo {author} {\bibfnamefont {B.}~\bibnamefont
  {Lavraud}}\ and\ \bibinfo {author} {\bibfnamefont {J.~E.}\ \bibnamefont
  {Borovsky}},\ }\bibfield  {title} {\enquote {\bibinfo {title} {Altered solar
  wind-magnetosphere interaction at low mach numbers: Coronal mass
  ejections},}\ }\href {https://doi.org/10.1029/2008JA013192} {\bibfield
  {journal} {\bibinfo  {journal} {Journal of Geophysical Research: Space
  Physics}\ }\textbf {\bibinfo {volume} {113}} (\bibinfo {year} {2008}),\
  10.1029/2008JA013192}\BibitemShut {NoStop}%
\bibitem [{\citenamefont {Moore}, \citenamefont {Nykyri},\ and\ \citenamefont
  {Dimmock}(2016)}]{Moore_NPhys2016}%
  \BibitemOpen
  \bibfield  {author} {\bibinfo {author} {\bibfnamefont {T.~W.}\ \bibnamefont
  {Moore}}, \bibinfo {author} {\bibfnamefont {K.}~\bibnamefont {Nykyri}},\ and\
  \bibinfo {author} {\bibfnamefont {A.~P.}\ \bibnamefont {Dimmock}},\
  }\bibfield  {title} {\enquote {\bibinfo {title} {Cross-scale energy transport
  in space plasmas},}\ }\href {https://doi.org/10.1038/nphys3869} {\bibfield
  {journal} {\bibinfo  {journal} {Nat Phys}\ }\textbf {\bibinfo {volume}
  {12}},\ \bibinfo {pages} {1164--1169} (\bibinfo {year} {2016})}\BibitemShut
  {NoStop}%
\bibitem [{\citenamefont {Liao}\ \emph {et~al.}(2015)\citenamefont {Liao},
  \citenamefont {Li}, \citenamefont {Hartigan}, \citenamefont {Graham},
  \citenamefont {Fiksel}, \citenamefont {Frank}, \citenamefont {Foster},\ and\
  \citenamefont {Kuranz}}]{Liao_HEDP2015}%
  \BibitemOpen
  \bibfield  {author} {\bibinfo {author} {\bibfnamefont {A.~S.}\ \bibnamefont
  {Liao}}, \bibinfo {author} {\bibfnamefont {S.}~\bibnamefont {Li}}, \bibinfo
  {author} {\bibfnamefont {P.}~\bibnamefont {Hartigan}}, \bibinfo {author}
  {\bibfnamefont {P.}~\bibnamefont {Graham}}, \bibinfo {author} {\bibfnamefont
  {G.}~\bibnamefont {Fiksel}}, \bibinfo {author} {\bibfnamefont
  {A.}~\bibnamefont {Frank}}, \bibinfo {author} {\bibfnamefont
  {J.}~\bibnamefont {Foster}},\ and\ \bibinfo {author} {\bibfnamefont
  {C.}~\bibnamefont {Kuranz}},\ }\bibfield  {title} {\enquote {\bibinfo {title}
  {Numerical simulation of an experimental analogue of a planetary
  magnetosphere},}\ }\href {https://doi.org/10.1016/j.hedp.2014.09.005}
  {\bibfield  {journal} {\bibinfo  {journal} {High Energy Density Physics}\
  }\textbf {\bibinfo {volume} {17, Part A}},\ \bibinfo {pages} {38--41}
  (\bibinfo {year} {2015})}\BibitemShut {NoStop}%
\bibitem [{\citenamefont {Froula}\ \emph {et~al.}(2011)\citenamefont {Froula},
  \citenamefont {Glenzer}, \citenamefont {Luhmann},\ and\ \citenamefont
  {Sheffield}}]{FroulaBook}%
  \BibitemOpen
  \bibinfo {editor} {\bibfnamefont {D.~H.}\ \bibnamefont {Froula}}, \bibinfo
  {editor} {\bibfnamefont {S.~H.}\ \bibnamefont {Glenzer}}, \bibinfo {editor}
  {\bibfnamefont {N.~C.}\ \bibnamefont {Luhmann}},\ and\ \bibinfo {editor}
  {\bibfnamefont {J.}~\bibnamefont {Sheffield}},\ eds.,\ \href@noop {} {\emph
  {\bibinfo {title} {Plasma Scattering of Electromagnetic Radiation}}},\
  \bibinfo {edition} {second edition}\ ed.\ (\bibinfo  {publisher} {Academic
  Press},\ \bibinfo {address} {Boston},\ \bibinfo {year} {2011})\BibitemShut
  {NoStop}%
\bibitem [{\citenamefont {Froula}\ \emph {et~al.}(2006)\citenamefont {Froula},
  \citenamefont {Ross}, \citenamefont {Divol},\ and\ \citenamefont
  {Glenzer}}]{Froula_RSI2006}%
  \BibitemOpen
  \bibfield  {author} {\bibinfo {author} {\bibfnamefont {D.~H.}\ \bibnamefont
  {Froula}}, \bibinfo {author} {\bibfnamefont {J.~S.}\ \bibnamefont {Ross}},
  \bibinfo {author} {\bibfnamefont {L.}~\bibnamefont {Divol}},\ and\ \bibinfo
  {author} {\bibfnamefont {S.~H.}\ \bibnamefont {Glenzer}},\ }\bibfield
  {title} {\enquote {\bibinfo {title} {Thomson-scattering techniques to
  diagnose local electron and ion temperatures, density, and plasma wave
  amplitudes in laser produced plasmas (invited)},}\ }\href
  {https://doi.org/10.1063/1.2336451} {\bibfield  {journal} {\bibinfo
  {journal} {Review of Scientific Instruments}\ }\textbf {\bibinfo {volume}
  {77}},\ \bibinfo {pages} {10E522} (\bibinfo {year} {2006})}\BibitemShut
  {NoStop}%
\bibitem [{\citenamefont {Katz}\ \emph {et~al.}(2013)\citenamefont {Katz},
  \citenamefont {Ross}, \citenamefont {Sorce},\ and\ \citenamefont
  {Froula}}]{Katz_JInst2013}%
  \BibitemOpen
  \bibfield  {author} {\bibinfo {author} {\bibfnamefont {J.}~\bibnamefont
  {Katz}}, \bibinfo {author} {\bibfnamefont {J.~S.}\ \bibnamefont {Ross}},
  \bibinfo {author} {\bibfnamefont {C.}~\bibnamefont {Sorce}},\ and\ \bibinfo
  {author} {\bibfnamefont {D.~H.}\ \bibnamefont {Froula}},\ }\bibfield  {title}
  {\enquote {\bibinfo {title} {A reflective image-rotating periscope for
  spatially resolved thomson-scattering experiments on omega},}\ }\href
  {https://doi.org/10.1088/1748-0221/8/12/C12009} {\bibfield  {journal}
  {\bibinfo  {journal} {Journal of Instrumentation}\ }\textbf {\bibinfo
  {volume} {8}},\ \bibinfo {pages} {C12009} (\bibinfo {year}
  {2013})}\BibitemShut {NoStop}%
\bibitem [{\citenamefont {Li}\ \emph {et~al.}(2006)\citenamefont {Li},
  \citenamefont {S{\'e}guin}, \citenamefont {Frenje}, \citenamefont {Rygg},
  \citenamefont {Petrasso}, \citenamefont {Town}, \citenamefont {Amendt},
  \citenamefont {Hatchett}, \citenamefont {Landen}, \citenamefont {Mackinnon},
  \citenamefont {Patel}, \citenamefont {Smalyuk}, \citenamefont {Knauer},
  \citenamefont {Sangster},\ and\ \citenamefont {Stoeckl}}]{Li_RSI2006}%
  \BibitemOpen
  \bibfield  {author} {\bibinfo {author} {\bibfnamefont {C.~K.}\ \bibnamefont
  {Li}}, \bibinfo {author} {\bibfnamefont {F.~H.}\ \bibnamefont {S{\'e}guin}},
  \bibinfo {author} {\bibfnamefont {J.~A.}\ \bibnamefont {Frenje}}, \bibinfo
  {author} {\bibfnamefont {J.~R.}\ \bibnamefont {Rygg}}, \bibinfo {author}
  {\bibfnamefont {R.~D.}\ \bibnamefont {Petrasso}}, \bibinfo {author}
  {\bibfnamefont {R.~P.~J.}\ \bibnamefont {Town}}, \bibinfo {author}
  {\bibfnamefont {P.~A.}\ \bibnamefont {Amendt}}, \bibinfo {author}
  {\bibfnamefont {S.~P.}\ \bibnamefont {Hatchett}}, \bibinfo {author}
  {\bibfnamefont {O.~L.}\ \bibnamefont {Landen}}, \bibinfo {author}
  {\bibfnamefont {A.~J.}\ \bibnamefont {Mackinnon}}, \bibinfo {author}
  {\bibfnamefont {P.~K.}\ \bibnamefont {Patel}}, \bibinfo {author}
  {\bibfnamefont {V.~A.}\ \bibnamefont {Smalyuk}}, \bibinfo {author}
  {\bibfnamefont {J.~P.}\ \bibnamefont {Knauer}}, \bibinfo {author}
  {\bibfnamefont {T.~C.}\ \bibnamefont {Sangster}},\ and\ \bibinfo {author}
  {\bibfnamefont {C.}~\bibnamefont {Stoeckl}},\ }\bibfield  {title} {\enquote
  {\bibinfo {title} {Monoenergetic proton backlighter for measuring e and b
  fields and for radiographing implosions and high-energy density plasmas
  (invited)},}\ }\href {https://doi.org/10.1063/1.2228252} {\bibfield
  {journal} {\bibinfo  {journal} {Review of Scientific Instruments}\ }\textbf
  {\bibinfo {volume} {77}},\ \bibinfo {pages} {10E725} (\bibinfo {year}
  {2006})}\BibitemShut {NoStop}%
\bibitem [{\citenamefont {Li}\ \emph {et~al.}(2009)\citenamefont {Li},
  \citenamefont {S{\'e}guin}, \citenamefont {Frenje}, \citenamefont {Manuel},
  \citenamefont {Casey}, \citenamefont {Sinenian}, \citenamefont {Petrasso},
  \citenamefont {Amendt}, \citenamefont {Landen}, \citenamefont {Rygg},
  \citenamefont {Town}, \citenamefont {Betti}, \citenamefont {Delettrez},
  \citenamefont {Knauer}, \citenamefont {Marshall}, \citenamefont {Meyerhofer},
  \citenamefont {Sangster}, \citenamefont {Shvarts}, \citenamefont {Smalyuk},
  \citenamefont {Soures}, \citenamefont {Back}, \citenamefont {Kilkenny},\ and\
  \citenamefont {Nikroo}}]{Li_PoP2009}%
  \BibitemOpen
  \bibfield  {author} {\bibinfo {author} {\bibfnamefont {C.~K.}\ \bibnamefont
  {Li}}, \bibinfo {author} {\bibfnamefont {F.~H.}\ \bibnamefont {S{\'e}guin}},
  \bibinfo {author} {\bibfnamefont {J.~A.}\ \bibnamefont {Frenje}}, \bibinfo
  {author} {\bibfnamefont {M.}~\bibnamefont {Manuel}}, \bibinfo {author}
  {\bibfnamefont {D.}~\bibnamefont {Casey}}, \bibinfo {author} {\bibfnamefont
  {N.}~\bibnamefont {Sinenian}}, \bibinfo {author} {\bibfnamefont {R.~D.}\
  \bibnamefont {Petrasso}}, \bibinfo {author} {\bibfnamefont {P.~A.}\
  \bibnamefont {Amendt}}, \bibinfo {author} {\bibfnamefont {O.~L.}\
  \bibnamefont {Landen}}, \bibinfo {author} {\bibfnamefont {J.~R.}\
  \bibnamefont {Rygg}}, \bibinfo {author} {\bibfnamefont {R.~P.~J.}\
  \bibnamefont {Town}}, \bibinfo {author} {\bibfnamefont {R.}~\bibnamefont
  {Betti}}, \bibinfo {author} {\bibfnamefont {J.}~\bibnamefont {Delettrez}},
  \bibinfo {author} {\bibfnamefont {J.~P.}\ \bibnamefont {Knauer}}, \bibinfo
  {author} {\bibfnamefont {F.}~\bibnamefont {Marshall}}, \bibinfo {author}
  {\bibfnamefont {D.~D.}\ \bibnamefont {Meyerhofer}}, \bibinfo {author}
  {\bibfnamefont {T.~C.}\ \bibnamefont {Sangster}}, \bibinfo {author}
  {\bibfnamefont {D.}~\bibnamefont {Shvarts}}, \bibinfo {author} {\bibfnamefont
  {V.~A.}\ \bibnamefont {Smalyuk}}, \bibinfo {author} {\bibfnamefont {J.~M.}\
  \bibnamefont {Soures}}, \bibinfo {author} {\bibfnamefont {C.~A.}\
  \bibnamefont {Back}}, \bibinfo {author} {\bibfnamefont {J.~D.}\ \bibnamefont
  {Kilkenny}},\ and\ \bibinfo {author} {\bibfnamefont {A.}~\bibnamefont
  {Nikroo}},\ }\bibfield  {title} {\enquote {\bibinfo {title} {Proton
  radiography of dynamic electric and magnetic fields in laser-produced
  high-energy-density plasmas},}\ }\href {https://doi.org/10.1063/1.3096781}
  {\bibfield  {journal} {\bibinfo  {journal} {Physics of Plasmas}\ }\textbf
  {\bibinfo {volume} {16}},\ \bibinfo {pages} {056304} (\bibinfo {year}
  {2009})}\BibitemShut {NoStop}%
\bibitem [{\citenamefont {Levesque}\ and\ \citenamefont
  {Beesley}(2021)}]{Levesque_RSI2021}%
  \BibitemOpen
  \bibfield  {author} {\bibinfo {author} {\bibfnamefont {J.~M.}\ \bibnamefont
  {Levesque}}\ and\ \bibinfo {author} {\bibfnamefont {L.~J.}\ \bibnamefont
  {Beesley}},\ }\bibfield  {title} {\enquote {\bibinfo {title} {Reconstructing
  magnetic deflections from sets of proton images using differential
  evolution},}\ }\href {https://doi.org/10.1063/5.0054862} {\bibfield
  {journal} {\bibinfo  {journal} {Review of Scientific Instruments}\ }\textbf
  {\bibinfo {volume} {92}},\ \bibinfo {pages} {093505} (\bibinfo {year}
  {2021})}\BibitemShut {NoStop}%
\bibitem [{\citenamefont {Burdiak}\ \emph {et~al.}(2017)\citenamefont
  {Burdiak}, \citenamefont {Lebedev}, \citenamefont {Bland}, \citenamefont
  {Clayson}, \citenamefont {Hare}, \citenamefont {Suttle}, \citenamefont
  {Suzuki-Vidal}, \citenamefont {Garcia}, \citenamefont {Chittenden},
  \citenamefont {Bott-Suzuki}, \citenamefont {Ciardi}, \citenamefont {Frank},\
  and\ \citenamefont {Lane}}]{Burdiak_PoP2017}%
  \BibitemOpen
  \bibfield  {author} {\bibinfo {author} {\bibfnamefont {G.~C.}\ \bibnamefont
  {Burdiak}}, \bibinfo {author} {\bibfnamefont {S.~V.}\ \bibnamefont
  {Lebedev}}, \bibinfo {author} {\bibfnamefont {S.~N.}\ \bibnamefont {Bland}},
  \bibinfo {author} {\bibfnamefont {T.}~\bibnamefont {Clayson}}, \bibinfo
  {author} {\bibfnamefont {J.}~\bibnamefont {Hare}}, \bibinfo {author}
  {\bibfnamefont {L.}~\bibnamefont {Suttle}}, \bibinfo {author} {\bibfnamefont
  {F.}~\bibnamefont {Suzuki-Vidal}}, \bibinfo {author} {\bibfnamefont {D.~C.}\
  \bibnamefont {Garcia}}, \bibinfo {author} {\bibfnamefont {J.~P.}\
  \bibnamefont {Chittenden}}, \bibinfo {author} {\bibfnamefont
  {S.}~\bibnamefont {Bott-Suzuki}}, \bibinfo {author} {\bibfnamefont
  {A.}~\bibnamefont {Ciardi}}, \bibinfo {author} {\bibfnamefont
  {A.}~\bibnamefont {Frank}},\ and\ \bibinfo {author} {\bibfnamefont {T.~S.}\
  \bibnamefont {Lane}},\ }\bibfield  {title} {\enquote {\bibinfo {title} {The
  structure of bow shocks formed by the interaction of pulsed-power driven
  magnetised plasma flows with conducting obstacles},}\ }\href
  {https://doi.org/10.1063/1.4993187} {\bibfield  {journal} {\bibinfo
  {journal} {Physics of Plasmas}\ }\textbf {\bibinfo {volume} {24}},\ \bibinfo
  {pages} {072713} (\bibinfo {year} {2017})}\BibitemShut {NoStop}%
\bibitem [{\citenamefont {{Suttle}}\ \emph {et~al.}(2020)\citenamefont
  {{Suttle}}, \citenamefont {{Burdiak}}, \citenamefont {{Cheung}},
  \citenamefont {{Clayson}}, \citenamefont {{Halliday}}, \citenamefont
  {{Hare}}, \citenamefont {{Rusli}}, \citenamefont {{Russell}}, \citenamefont
  {{Tubman}}, \citenamefont {{Ciardi}}, \citenamefont {{Loureiro}},
  \citenamefont {{Li}}, \citenamefont {{Frank}},\ and\ \citenamefont
  {{Lebedev}}}]{Suttle_2020PPCF}%
  \BibitemOpen
  \bibfield  {author} {\bibinfo {author} {\bibfnamefont {L.~G.}\ \bibnamefont
  {{Suttle}}}, \bibinfo {author} {\bibfnamefont {G.~C.}\ \bibnamefont
  {{Burdiak}}}, \bibinfo {author} {\bibfnamefont {C.~L.}\ \bibnamefont
  {{Cheung}}}, \bibinfo {author} {\bibfnamefont {T.}~\bibnamefont {{Clayson}}},
  \bibinfo {author} {\bibfnamefont {J.~W.~D.}\ \bibnamefont {{Halliday}}},
  \bibinfo {author} {\bibfnamefont {J.~D.}\ \bibnamefont {{Hare}}}, \bibinfo
  {author} {\bibfnamefont {S.}~\bibnamefont {{Rusli}}}, \bibinfo {author}
  {\bibfnamefont {D.~R.}\ \bibnamefont {{Russell}}}, \bibinfo {author}
  {\bibfnamefont {E.~R.}\ \bibnamefont {{Tubman}}}, \bibinfo {author}
  {\bibfnamefont {A.}~\bibnamefont {{Ciardi}}}, \bibinfo {author}
  {\bibfnamefont {N.~F.}\ \bibnamefont {{Loureiro}}}, \bibinfo {author}
  {\bibfnamefont {J.}~\bibnamefont {{Li}}}, \bibinfo {author} {\bibfnamefont
  {A.}~\bibnamefont {{Frank}}},\ and\ \bibinfo {author} {\bibfnamefont {S.~V.}\
  \bibnamefont {{Lebedev}}},\ }\bibfield  {title} {\enquote {\bibinfo {title}
  {{Interactions of magnetized plasma flows in pulsed-power driven
  experiments}},}\ }\href {https://doi.org/10.1088/1361-6587/ab5296} {\bibfield
   {journal} {\bibinfo  {journal} {Plasma Physics and Controlled Fusion}\
  }\textbf {\bibinfo {volume} {62}},\ \bibinfo {pages} {014020} (\bibinfo
  {year} {2020})}\BibitemShut {NoStop}%
\bibitem [{\citenamefont {Suttle}\ \emph {et~al.}(2021)\citenamefont {Suttle},
  \citenamefont {Hare}, \citenamefont {Halliday}, \citenamefont {Merlini},
  \citenamefont {Russell}, \citenamefont {Tubman}, \citenamefont
  {Valenzuela-Villaseca}, \citenamefont {Rozmus}, \citenamefont {Bruulsema},\
  and\ \citenamefont {Lebedev}}]{Suttle_RSI2021}%
  \BibitemOpen
  \bibfield  {author} {\bibinfo {author} {\bibfnamefont {L.~G.}\ \bibnamefont
  {Suttle}}, \bibinfo {author} {\bibfnamefont {J.~D.}\ \bibnamefont {Hare}},
  \bibinfo {author} {\bibfnamefont {J.~W.~D.}\ \bibnamefont {Halliday}},
  \bibinfo {author} {\bibfnamefont {S.}~\bibnamefont {Merlini}}, \bibinfo
  {author} {\bibfnamefont {D.~R.}\ \bibnamefont {Russell}}, \bibinfo {author}
  {\bibfnamefont {E.~R.}\ \bibnamefont {Tubman}}, \bibinfo {author}
  {\bibfnamefont {V.}~\bibnamefont {Valenzuela-Villaseca}}, \bibinfo {author}
  {\bibfnamefont {W.}~\bibnamefont {Rozmus}}, \bibinfo {author} {\bibfnamefont
  {C.}~\bibnamefont {Bruulsema}},\ and\ \bibinfo {author} {\bibfnamefont
  {S.~V.}\ \bibnamefont {Lebedev}},\ }\bibfield  {title} {\enquote {\bibinfo
  {title} {Collective optical thomson scattering in pulsed-power driven high
  energy density physics experiments (invited)},}\ }\href
  {https://doi.org/10.1063/5.0041118} {\bibfield  {journal} {\bibinfo
  {journal} {Review of Scientific Instruments}\ }\textbf {\bibinfo {volume}
  {92}},\ \bibinfo {pages} {033542} (\bibinfo {year} {2021})}\BibitemShut
  {NoStop}%
\bibitem [{\citenamefont {Rigby}\ \emph {et~al.}(2018)\citenamefont {Rigby},
  \citenamefont {Cruz}, \citenamefont {Albertazzi}, \citenamefont {Bamford},
  \citenamefont {Bell}, \citenamefont {Cross}, \citenamefont {Fraschetti},
  \citenamefont {Graham}, \citenamefont {Hara}, \citenamefont {Kozlowski},
  \citenamefont {Kuramitsu}, \citenamefont {Lamb}, \citenamefont {Lebedev},
  \citenamefont {Marques}, \citenamefont {Miniati}, \citenamefont {Morita},
  \citenamefont {Oliver}, \citenamefont {Reville}, \citenamefont {Sakawa},
  \citenamefont {Sarkar}, \citenamefont {Spindloe}, \citenamefont {Trines},
  \citenamefont {Tzeferacos}, \citenamefont {Silva}, \citenamefont {Bingham},
  \citenamefont {Koenig},\ and\ \citenamefont {Gregori}}]{Rigby_NatPhys2018}%
  \BibitemOpen
  \bibfield  {author} {\bibinfo {author} {\bibfnamefont {A.}~\bibnamefont
  {Rigby}}, \bibinfo {author} {\bibfnamefont {F.}~\bibnamefont {Cruz}},
  \bibinfo {author} {\bibfnamefont {B.}~\bibnamefont {Albertazzi}}, \bibinfo
  {author} {\bibfnamefont {R.}~\bibnamefont {Bamford}}, \bibinfo {author}
  {\bibfnamefont {A.~R.}\ \bibnamefont {Bell}}, \bibinfo {author}
  {\bibfnamefont {J.~E.}\ \bibnamefont {Cross}}, \bibinfo {author}
  {\bibfnamefont {F.}~\bibnamefont {Fraschetti}}, \bibinfo {author}
  {\bibfnamefont {P.}~\bibnamefont {Graham}}, \bibinfo {author} {\bibfnamefont
  {Y.}~\bibnamefont {Hara}}, \bibinfo {author} {\bibfnamefont {P.~M.}\
  \bibnamefont {Kozlowski}}, \bibinfo {author} {\bibfnamefont {Y.}~\bibnamefont
  {Kuramitsu}}, \bibinfo {author} {\bibfnamefont {D.~Q.}\ \bibnamefont {Lamb}},
  \bibinfo {author} {\bibfnamefont {S.}~\bibnamefont {Lebedev}}, \bibinfo
  {author} {\bibfnamefont {J.~R.}\ \bibnamefont {Marques}}, \bibinfo {author}
  {\bibfnamefont {F.}~\bibnamefont {Miniati}}, \bibinfo {author} {\bibfnamefont
  {T.}~\bibnamefont {Morita}}, \bibinfo {author} {\bibfnamefont
  {M.}~\bibnamefont {Oliver}}, \bibinfo {author} {\bibfnamefont
  {B.}~\bibnamefont {Reville}}, \bibinfo {author} {\bibfnamefont
  {Y.}~\bibnamefont {Sakawa}}, \bibinfo {author} {\bibfnamefont
  {S.}~\bibnamefont {Sarkar}}, \bibinfo {author} {\bibfnamefont
  {C.}~\bibnamefont {Spindloe}}, \bibinfo {author} {\bibfnamefont
  {R.}~\bibnamefont {Trines}}, \bibinfo {author} {\bibfnamefont
  {P.}~\bibnamefont {Tzeferacos}}, \bibinfo {author} {\bibfnamefont {L.~O.}\
  \bibnamefont {Silva}}, \bibinfo {author} {\bibfnamefont {R.}~\bibnamefont
  {Bingham}}, \bibinfo {author} {\bibfnamefont {M.}~\bibnamefont {Koenig}},\
  and\ \bibinfo {author} {\bibfnamefont {G.}~\bibnamefont {Gregori}},\
  }\bibfield  {title} {\enquote {\bibinfo {title} {Electron acceleration by
  wave turbulence in a magnetized plasma},}\ }\href
  {https://doi.org/10.1038/s41567-018-0059-2} {\bibfield  {journal} {\bibinfo
  {journal} {Nature Physics}\ }\textbf {\bibinfo {volume} {14}},\ \bibinfo
  {pages} {475--479} (\bibinfo {year} {2018})}\BibitemShut {NoStop}%
\bibitem [{\citenamefont {Fiksel}\ \emph {et~al.}(2015)\citenamefont {Fiksel},
  \citenamefont {Agliata}, \citenamefont {Barnak}, \citenamefont {Brent},
  \citenamefont {Chang}, \citenamefont {Folnsbee}, \citenamefont {Gates},
  \citenamefont {Hasset}, \citenamefont {Lonobile}, \citenamefont {Magoon},
  \citenamefont {Mastrosimone}, \citenamefont {Shoup},\ and\ \citenamefont
  {Betti}}]{Fiksel_RSI2015}%
  \BibitemOpen
  \bibfield  {author} {\bibinfo {author} {\bibfnamefont {G.}~\bibnamefont
  {Fiksel}}, \bibinfo {author} {\bibfnamefont {A.}~\bibnamefont {Agliata}},
  \bibinfo {author} {\bibfnamefont {D.}~\bibnamefont {Barnak}}, \bibinfo
  {author} {\bibfnamefont {G.}~\bibnamefont {Brent}}, \bibinfo {author}
  {\bibfnamefont {P.~Y.}\ \bibnamefont {Chang}}, \bibinfo {author}
  {\bibfnamefont {L.}~\bibnamefont {Folnsbee}}, \bibinfo {author}
  {\bibfnamefont {G.}~\bibnamefont {Gates}}, \bibinfo {author} {\bibfnamefont
  {D.}~\bibnamefont {Hasset}}, \bibinfo {author} {\bibfnamefont
  {D.}~\bibnamefont {Lonobile}}, \bibinfo {author} {\bibfnamefont
  {J.}~\bibnamefont {Magoon}}, \bibinfo {author} {\bibfnamefont
  {D.}~\bibnamefont {Mastrosimone}}, \bibinfo {author} {\bibfnamefont {M.~J.}\
  \bibnamefont {Shoup}},\ and\ \bibinfo {author} {\bibfnamefont
  {R.}~\bibnamefont {Betti}},\ }\bibfield  {title} {\enquote {\bibinfo {title}
  {Note: Experimental platform for magnetized high-energy-density plasma
  studies at the omega laser facility},}\ }\href
  {https://doi.org/10.1063/1.4905625} {\bibfield  {journal} {\bibinfo
  {journal} {Review of Scientific Instruments}\ }\textbf {\bibinfo {volume}
  {86}},\ \bibinfo {pages} {016105} (\bibinfo {year} {2015})}\BibitemShut
  {NoStop}%
\bibitem [{\citenamefont {Lee}\ and\ \citenamefont
  {Deane}(2009)}]{Lee_JCompPhys2009}%
  \BibitemOpen
  \bibfield  {author} {\bibinfo {author} {\bibfnamefont {D.}~\bibnamefont
  {Lee}}\ and\ \bibinfo {author} {\bibfnamefont {A.~E.}\ \bibnamefont
  {Deane}},\ }\bibfield  {title} {\enquote {\bibinfo {title} {An unsplit
  staggered mesh scheme for multidimensional magnetohydrodynamics},}\ }\href
  {https://doi.org/https://doi.org/10.1016/j.jcp.2008.08.026} {\bibfield
  {journal} {\bibinfo  {journal} {Journal of Computational Physics}\ }\textbf
  {\bibinfo {volume} {228}},\ \bibinfo {pages} {952--975} (\bibinfo {year}
  {2009})}\BibitemShut {NoStop}%
\bibitem [{\citenamefont {Tzeferacos}\ \emph {et~al.}(2012)\citenamefont
  {Tzeferacos}, \citenamefont {Fatenejad}, \citenamefont {Flocke},
  \citenamefont {Gregori}, \citenamefont {Lamb}, \citenamefont {Lee},
  \citenamefont {Meinecke}, \citenamefont {Scopatz},\ and\ \citenamefont
  {Weide}}]{Tzeferacos_HEDP2012}%
  \BibitemOpen
  \bibfield  {author} {\bibinfo {author} {\bibfnamefont {P.}~\bibnamefont
  {Tzeferacos}}, \bibinfo {author} {\bibfnamefont {M.}~\bibnamefont
  {Fatenejad}}, \bibinfo {author} {\bibfnamefont {N.}~\bibnamefont {Flocke}},
  \bibinfo {author} {\bibfnamefont {G.}~\bibnamefont {Gregori}}, \bibinfo
  {author} {\bibfnamefont {D.}~\bibnamefont {Lamb}}, \bibinfo {author}
  {\bibfnamefont {D.}~\bibnamefont {Lee}}, \bibinfo {author} {\bibfnamefont
  {J.}~\bibnamefont {Meinecke}}, \bibinfo {author} {\bibfnamefont
  {A.}~\bibnamefont {Scopatz}},\ and\ \bibinfo {author} {\bibfnamefont
  {K.}~\bibnamefont {Weide}},\ }\bibfield  {title} {\enquote {\bibinfo {title}
  {Flash magnetohydrodynamic simulations of shock-generated magnetic field
  experiments},}\ }\href
  {https://doi.org/https://doi.org/10.1016/j.hedp.2012.08.001} {\bibfield
  {journal} {\bibinfo  {journal} {High Energy Density Physics}\ }\textbf
  {\bibinfo {volume} {8}},\ \bibinfo {pages} {322--328} (\bibinfo {year}
  {2012})}\BibitemShut {NoStop}%
\bibitem [{\citenamefont {Chapman}\ and\ \citenamefont
  {Ferraro}(1931)}]{Chapman_Ferraro_1931b}%
  \BibitemOpen
  \bibfield  {author} {\bibinfo {author} {\bibfnamefont {S.}~\bibnamefont
  {Chapman}}\ and\ \bibinfo {author} {\bibfnamefont {V.~C.~A.}\ \bibnamefont
  {Ferraro}},\ }\bibfield  {title} {\enquote {\bibinfo {title} {A new theory of
  magnetic storms},}\ }\href {https://doi.org/10.1029/TE036i003p00171}
  {\bibfield  {journal} {\bibinfo  {journal} {Terrestrial Magnetism and
  Atmospheric Electricity}\ }\textbf {\bibinfo {volume} {36}},\ \bibinfo
  {pages} {171--186} (\bibinfo {year} {1931})}\BibitemShut {NoStop}%
\bibitem [{\citenamefont {Ross}\ \emph {et~al.}(2006)\citenamefont {Ross},
  \citenamefont {Froula}, \citenamefont {Mackinnon}, \citenamefont {Sorce},
  \citenamefont {Meezan}, \citenamefont {Glenzer}, \citenamefont {Armstrong},
  \citenamefont {Bahr}, \citenamefont {Huff},\ and\ \citenamefont
  {Thorp}}]{Ross_RSI2006}%
  \BibitemOpen
  \bibfield  {author} {\bibinfo {author} {\bibfnamefont {J.~S.}\ \bibnamefont
  {Ross}}, \bibinfo {author} {\bibfnamefont {D.~H.}\ \bibnamefont {Froula}},
  \bibinfo {author} {\bibfnamefont {A.~J.}\ \bibnamefont {Mackinnon}}, \bibinfo
  {author} {\bibfnamefont {C.}~\bibnamefont {Sorce}}, \bibinfo {author}
  {\bibfnamefont {N.}~\bibnamefont {Meezan}}, \bibinfo {author} {\bibfnamefont
  {S.~H.}\ \bibnamefont {Glenzer}}, \bibinfo {author} {\bibfnamefont
  {W.}~\bibnamefont {Armstrong}}, \bibinfo {author} {\bibfnamefont
  {R.}~\bibnamefont {Bahr}}, \bibinfo {author} {\bibfnamefont {R.}~\bibnamefont
  {Huff}},\ and\ \bibinfo {author} {\bibfnamefont {K.}~\bibnamefont {Thorp}},\
  }\bibfield  {title} {\enquote {\bibinfo {title} {Implementation of imaging
  thomson scattering on the omega laser},}\ }\href
  {https://doi.org/10.1063/1.2220077} {\bibfield  {journal} {\bibinfo
  {journal} {Review of Scientific Instruments}\ }\textbf {\bibinfo {volume}
  {77}},\ \bibinfo {pages} {10E520} (\bibinfo {year} {2006})}\BibitemShut
  {NoStop}%
\bibitem [{\citenamefont {Ross}\ \emph {et~al.}(2010)\citenamefont {Ross},
  \citenamefont {Glenzer}, \citenamefont {Palastro}, \citenamefont {Pollock},
  \citenamefont {Price}, \citenamefont {Tynan},\ and\ \citenamefont
  {Froula}}]{Ross_RSI2010}%
  \BibitemOpen
  \bibfield  {author} {\bibinfo {author} {\bibfnamefont {J.~S.}\ \bibnamefont
  {Ross}}, \bibinfo {author} {\bibfnamefont {S.~H.}\ \bibnamefont {Glenzer}},
  \bibinfo {author} {\bibfnamefont {J.~P.}\ \bibnamefont {Palastro}}, \bibinfo
  {author} {\bibfnamefont {B.~B.}\ \bibnamefont {Pollock}}, \bibinfo {author}
  {\bibfnamefont {D.}~\bibnamefont {Price}}, \bibinfo {author} {\bibfnamefont
  {G.~R.}\ \bibnamefont {Tynan}},\ and\ \bibinfo {author} {\bibfnamefont
  {D.~H.}\ \bibnamefont {Froula}},\ }\bibfield  {title} {\enquote {\bibinfo
  {title} {Thomson-scattering measurements in the collective and noncollective
  regimes in laser produced plasmas (invited)},}\ }\href
  {https://doi.org/10.1063/1.3478975} {\bibfield  {journal} {\bibinfo
  {journal} {Review of Scientific Instruments}\ }\textbf {\bibinfo {volume}
  {81}},\ \bibinfo {pages} {10D523} (\bibinfo {year} {2010})}\BibitemShut
  {NoStop}%
\bibitem [{\citenamefont {Follett}\ \emph {et~al.}(2016)\citenamefont
  {Follett}, \citenamefont {Delettrez}, \citenamefont {Edgell}, \citenamefont
  {Henchen}, \citenamefont {Katz}, \citenamefont {Myatt},\ and\ \citenamefont
  {Froula}}]{Follett_RSI2016}%
  \BibitemOpen
  \bibfield  {author} {\bibinfo {author} {\bibfnamefont {R.~K.}\ \bibnamefont
  {Follett}}, \bibinfo {author} {\bibfnamefont {J.~A.}\ \bibnamefont
  {Delettrez}}, \bibinfo {author} {\bibfnamefont {D.~H.}\ \bibnamefont
  {Edgell}}, \bibinfo {author} {\bibfnamefont {R.~J.}\ \bibnamefont {Henchen}},
  \bibinfo {author} {\bibfnamefont {J.}~\bibnamefont {Katz}}, \bibinfo {author}
  {\bibfnamefont {J.~F.}\ \bibnamefont {Myatt}},\ and\ \bibinfo {author}
  {\bibfnamefont {D.~H.}\ \bibnamefont {Froula}},\ }\bibfield  {title}
  {\enquote {\bibinfo {title} {Plasma characterization using ultraviolet
  thomson scattering from ion-acoustic and electron plasma waves (invited)},}\
  }\href {https://doi.org/10.1063/1.4959160} {\bibfield  {journal} {\bibinfo
  {journal} {Review of Scientific Instruments}\ }\textbf {\bibinfo {volume}
  {87}},\ \bibinfo {pages} {11E401} (\bibinfo {year} {2016})}\BibitemShut
  {NoStop}%
\bibitem [{Pri()}]{PrismSPECT_v6.5.0}%
  \BibitemOpen
  \href@noop {} {\emph {\bibinfo {title} {PrismSPECT version 6.5.0}}},\
  \bibinfo {organization} {Prism Computational Sciences, Inc.},\ \bibinfo
  {address} {www.prism-cs.com}\BibitemShut {NoStop}%
\bibitem [{\citenamefont {Priest}(1982)}]{PriestBook}%
  \BibitemOpen
  \bibfield  {author} {\bibinfo {author} {\bibfnamefont {E.~R.}\ \bibnamefont
  {Priest}},\ }\href@noop {} {\emph {\bibinfo {title} {Solar
  Magnetohydrodynamics}}}\ (\bibinfo  {publisher} {Springer, Dordrecht},\
  \bibinfo {year} {1982})\BibitemShut {NoStop}%
\bibitem [{\citenamefont {Draine}\ and\ \citenamefont
  {McKee}(1993)}]{Draine_AnnuRevAA1993}%
  \BibitemOpen
  \bibfield  {author} {\bibinfo {author} {\bibfnamefont {B.~T.}\ \bibnamefont
  {Draine}}\ and\ \bibinfo {author} {\bibfnamefont {C.~F.}\ \bibnamefont
  {McKee}},\ }\bibfield  {title} {\enquote {\bibinfo {title} {Theory of
  interstellar shocks},}\ }\href
  {https://doi.org/10.1146/annurev.aa.31.090193.002105} {\bibfield  {journal}
  {\bibinfo  {journal} {Annual Review of Astronomy and Astrophysics}\ }\textbf
  {\bibinfo {volume} {31}},\ \bibinfo {pages} {373--432} (\bibinfo {year}
  {1993})}\BibitemShut {NoStop}%
\bibitem [{\citenamefont {Hartigan}(2003)}]{Hartigan2003}%
  \BibitemOpen
  \bibfield  {author} {\bibinfo {author} {\bibfnamefont {P.}~\bibnamefont
  {Hartigan}},\ }\bibfield  {title} {\enquote {\bibinfo {title} {Shock waves in
  outflows from young stars},}\ }\href
  {https://doi.org/10.1023/B:ASTR.0000006209.56314.c8} {\bibfield  {journal}
  {\bibinfo  {journal} {Astrophysics and Space Science}\ }\textbf {\bibinfo
  {volume} {287}},\ \bibinfo {pages} {111--122} (\bibinfo {year}
  {2003})}\BibitemShut {NoStop}%
\bibitem [{\citenamefont {Gotchev}\ \emph {et~al.}(2009)\citenamefont
  {Gotchev}, \citenamefont {Knauer}, \citenamefont {Chang}, \citenamefont
  {Jang}, \citenamefont {Shoup}, \citenamefont {Meyerhofer},\ and\
  \citenamefont {Betti}}]{Gotchev_RSI2009}%
  \BibitemOpen
  \bibfield  {author} {\bibinfo {author} {\bibfnamefont {O.~V.}\ \bibnamefont
  {Gotchev}}, \bibinfo {author} {\bibfnamefont {J.~P.}\ \bibnamefont {Knauer}},
  \bibinfo {author} {\bibfnamefont {P.~Y.}\ \bibnamefont {Chang}}, \bibinfo
  {author} {\bibfnamefont {N.~W.}\ \bibnamefont {Jang}}, \bibinfo {author}
  {\bibfnamefont {M.~J.}\ \bibnamefont {Shoup}}, \bibinfo {author}
  {\bibfnamefont {D.~D.}\ \bibnamefont {Meyerhofer}},\ and\ \bibinfo {author}
  {\bibfnamefont {R.}~\bibnamefont {Betti}},\ }\bibfield  {title} {\enquote
  {\bibinfo {title} {Seeding magnetic fields for laser-driven flux compression
  in high-energy-density plasmas},}\ }\href {https://doi.org/10.1063/1.3115983}
  {\bibfield  {journal} {\bibinfo  {journal} {Review of Scientific
  Instruments}\ }\textbf {\bibinfo {volume} {80}},\ \bibinfo {pages} {043504}
  (\bibinfo {year} {2009})}\BibitemShut {NoStop}%
\bibitem [{\citenamefont {Kugland}\ \emph {et~al.}(2012)\citenamefont
  {Kugland}, \citenamefont {Ryutov}, \citenamefont {Plechaty}, \citenamefont
  {Ross},\ and\ \citenamefont {Park}}]{Kugland_RSI2012}%
  \BibitemOpen
  \bibfield  {author} {\bibinfo {author} {\bibfnamefont {N.~L.}\ \bibnamefont
  {Kugland}}, \bibinfo {author} {\bibfnamefont {D.~D.}\ \bibnamefont {Ryutov}},
  \bibinfo {author} {\bibfnamefont {C.}~\bibnamefont {Plechaty}}, \bibinfo
  {author} {\bibfnamefont {J.~S.}\ \bibnamefont {Ross}},\ and\ \bibinfo
  {author} {\bibfnamefont {H.-S.}\ \bibnamefont {Park}},\ }\bibfield  {title}
  {\enquote {\bibinfo {title} {Invited article: Relation between electric and
  magnetic field structures and their proton-beam images},}\ }\href
  {https://doi.org/10.1063/1.4750234} {\bibfield  {journal} {\bibinfo
  {journal} {Review of Scientific Instruments}\ }\textbf {\bibinfo {volume}
  {83}} (\bibinfo {year} {2012}),\ 10.1063/1.4750234}\BibitemShut {NoStop}%
\bibitem [{\citenamefont {Graziani}\ \emph {et~al.}(2017)\citenamefont
  {Graziani}, \citenamefont {Tzeferacos}, \citenamefont {Lamb},\ and\
  \citenamefont {Li}}]{Graziani_RSI2017}%
  \BibitemOpen
  \bibfield  {author} {\bibinfo {author} {\bibfnamefont {C.}~\bibnamefont
  {Graziani}}, \bibinfo {author} {\bibfnamefont {P.}~\bibnamefont
  {Tzeferacos}}, \bibinfo {author} {\bibfnamefont {D.~Q.}\ \bibnamefont
  {Lamb}},\ and\ \bibinfo {author} {\bibfnamefont {C.}~\bibnamefont {Li}},\
  }\bibfield  {title} {\enquote {\bibinfo {title} {Inferring morphology and
  strength of magnetic fields from proton radiographs},}\ }\href
  {https://doi.org/10.1063/1.5013029} {\bibfield  {journal} {\bibinfo
  {journal} {Review of Scientific Instruments}\ }\textbf {\bibinfo {volume}
  {88}},\ \bibinfo {pages} {123507} (\bibinfo {year} {2017})}\BibitemShut
  {NoStop}%
\bibitem [{\citenamefont {Bott}\ \emph {et~al.}(2017)\citenamefont {Bott},
  \citenamefont {Graziani}, \citenamefont {Tzeferacos}, \citenamefont {White},
  \citenamefont {Lamb}, \citenamefont {Gregori},\ and\ \citenamefont
  {Schekochihin}}]{Bott_JPP2017}%
  \BibitemOpen
  \bibfield  {author} {\bibinfo {author} {\bibfnamefont {A.~F.~A.}\
  \bibnamefont {Bott}}, \bibinfo {author} {\bibfnamefont {C.}~\bibnamefont
  {Graziani}}, \bibinfo {author} {\bibfnamefont {P.}~\bibnamefont
  {Tzeferacos}}, \bibinfo {author} {\bibfnamefont {T.~G.}\ \bibnamefont
  {White}}, \bibinfo {author} {\bibfnamefont {D.~Q.}\ \bibnamefont {Lamb}},
  \bibinfo {author} {\bibfnamefont {G.}~\bibnamefont {Gregori}},\ and\ \bibinfo
  {author} {\bibfnamefont {A.~A.}\ \bibnamefont {Schekochihin}},\ }\bibfield
  {title} {\enquote {\bibinfo {title} {Proton imaging of stochastic magnetic
  fields},}\ }\href {https://doi.org/10.1017/S0022377817000939} {\bibfield
  {journal} {\bibinfo  {journal} {Journal of Plasma Physics}\ }\textbf
  {\bibinfo {volume} {83}},\ \bibinfo {pages} {905830614} (\bibinfo {year}
  {2017})}\BibitemShut {NoStop}%
\bibitem [{\citenamefont {Chen}\ \emph {et~al.}(2017)\citenamefont {Chen},
  \citenamefont {Kasim}, \citenamefont {Ceurvorst}, \citenamefont {Ratan},
  \citenamefont {Sadler}, \citenamefont {Levy}, \citenamefont {Trines},
  \citenamefont {Bingham},\ and\ \citenamefont {Norreys}}]{Chen_PRE2017}%
  \BibitemOpen
  \bibfield  {author} {\bibinfo {author} {\bibfnamefont {N.~F.~Y.}\
  \bibnamefont {Chen}}, \bibinfo {author} {\bibfnamefont {M.~F.}\ \bibnamefont
  {Kasim}}, \bibinfo {author} {\bibfnamefont {L.}~\bibnamefont {Ceurvorst}},
  \bibinfo {author} {\bibfnamefont {N.}~\bibnamefont {Ratan}}, \bibinfo
  {author} {\bibfnamefont {J.}~\bibnamefont {Sadler}}, \bibinfo {author}
  {\bibfnamefont {M.~C.}\ \bibnamefont {Levy}}, \bibinfo {author}
  {\bibfnamefont {R.}~\bibnamefont {Trines}}, \bibinfo {author} {\bibfnamefont
  {R.}~\bibnamefont {Bingham}},\ and\ \bibinfo {author} {\bibfnamefont
  {P.}~\bibnamefont {Norreys}},\ }\bibfield  {title} {\enquote {\bibinfo
  {title} {Machine learning applied to proton radiography of
  high-energy-density plasmas},}\ }\href
  {https://doi.org/10.1103/PhysRevE.95.043305} {\bibfield  {journal} {\bibinfo
  {journal} {Phys. Rev. E}\ }\textbf {\bibinfo {volume} {95}},\ \bibinfo
  {pages} {043305} (\bibinfo {year} {2017})}\BibitemShut {NoStop}%
\bibitem [{\citenamefont {Kasim}\ \emph {et~al.}(2017)\citenamefont {Kasim},
  \citenamefont {Ceurvorst}, \citenamefont {Ratan}, \citenamefont {Sadler},
  \citenamefont {Chen}, \citenamefont {S\"avert}, \citenamefont {Trines},
  \citenamefont {Bingham}, \citenamefont {Burrows}, \citenamefont {Kaluza},\
  and\ \citenamefont {Norreys}}]{Kasim_PRE2017}%
  \BibitemOpen
  \bibfield  {author} {\bibinfo {author} {\bibfnamefont {M.~F.}\ \bibnamefont
  {Kasim}}, \bibinfo {author} {\bibfnamefont {L.}~\bibnamefont {Ceurvorst}},
  \bibinfo {author} {\bibfnamefont {N.}~\bibnamefont {Ratan}}, \bibinfo
  {author} {\bibfnamefont {J.}~\bibnamefont {Sadler}}, \bibinfo {author}
  {\bibfnamefont {N.}~\bibnamefont {Chen}}, \bibinfo {author} {\bibfnamefont
  {A.}~\bibnamefont {S\"avert}}, \bibinfo {author} {\bibfnamefont
  {R.}~\bibnamefont {Trines}}, \bibinfo {author} {\bibfnamefont
  {R.}~\bibnamefont {Bingham}}, \bibinfo {author} {\bibfnamefont {P.~N.}\
  \bibnamefont {Burrows}}, \bibinfo {author} {\bibfnamefont {M.~C.}\
  \bibnamefont {Kaluza}},\ and\ \bibinfo {author} {\bibfnamefont
  {P.}~\bibnamefont {Norreys}},\ }\bibfield  {title} {\enquote {\bibinfo
  {title} {Quantitative shadowgraphy and proton radiography for large intensity
  modulations},}\ }\href {https://doi.org/10.1103/PhysRevE.95.023306}
  {\bibfield  {journal} {\bibinfo  {journal} {Phys. Rev. E}\ }\textbf {\bibinfo
  {volume} {95}},\ \bibinfo {pages} {023306} (\bibinfo {year}
  {2017})}\BibitemShut {NoStop}%
\bibitem [{\citenamefont {Kasim}\ \emph {et~al.}(2019)\citenamefont {Kasim},
  \citenamefont {Bott}, \citenamefont {Tzeferacos}, \citenamefont {Lamb},
  \citenamefont {Gregori},\ and\ \citenamefont {Vinko}}]{Kasim_PRE2019}%
  \BibitemOpen
  \bibfield  {author} {\bibinfo {author} {\bibfnamefont {M.~F.}\ \bibnamefont
  {Kasim}}, \bibinfo {author} {\bibfnamefont {A.~F.~A.}\ \bibnamefont {Bott}},
  \bibinfo {author} {\bibfnamefont {P.}~\bibnamefont {Tzeferacos}}, \bibinfo
  {author} {\bibfnamefont {D.~Q.}\ \bibnamefont {Lamb}}, \bibinfo {author}
  {\bibfnamefont {G.}~\bibnamefont {Gregori}},\ and\ \bibinfo {author}
  {\bibfnamefont {S.~M.}\ \bibnamefont {Vinko}},\ }\bibfield  {title} {\enquote
  {\bibinfo {title} {Retrieving fields from proton radiography without source
  profiles},}\ }\href {https://doi.org/10.1103/PhysRevE.100.033208} {\bibfield
  {journal} {\bibinfo  {journal} {Phys. Rev. E}\ }\textbf {\bibinfo {volume}
  {100}},\ \bibinfo {pages} {033208} (\bibinfo {year} {2019})}\BibitemShut
  {NoStop}%
\bibitem [{\citenamefont {Hua}\ \emph {et~al.}(2019)\citenamefont {Hua},
  \citenamefont {Kim}, \citenamefont {Sherlock}, \citenamefont
  {Bailly-Grandvaux}, \citenamefont {Beg}, \citenamefont {McGuffey},
  \citenamefont {Wilks}, \citenamefont {Wen}, \citenamefont {Joglekar},
  \citenamefont {Mori},\ and\ \citenamefont {Ping}}]{Hua_PRL2019}%
  \BibitemOpen
  \bibfield  {author} {\bibinfo {author} {\bibfnamefont {R.}~\bibnamefont
  {Hua}}, \bibinfo {author} {\bibfnamefont {J.}~\bibnamefont {Kim}}, \bibinfo
  {author} {\bibfnamefont {M.}~\bibnamefont {Sherlock}}, \bibinfo {author}
  {\bibfnamefont {M.}~\bibnamefont {Bailly-Grandvaux}}, \bibinfo {author}
  {\bibfnamefont {F.~N.}\ \bibnamefont {Beg}}, \bibinfo {author} {\bibfnamefont
  {C.}~\bibnamefont {McGuffey}}, \bibinfo {author} {\bibfnamefont
  {S.}~\bibnamefont {Wilks}}, \bibinfo {author} {\bibfnamefont
  {H.}~\bibnamefont {Wen}}, \bibinfo {author} {\bibfnamefont {A.}~\bibnamefont
  {Joglekar}}, \bibinfo {author} {\bibfnamefont {W.}~\bibnamefont {Mori}},\
  and\ \bibinfo {author} {\bibfnamefont {Y.}~\bibnamefont {Ping}},\ }\bibfield
  {title} {\enquote {\bibinfo {title} {Self-generated magnetic and electric
  fields at a mach-6 shock front in a low density helium gas by dual-angle
  proton radiography},}\ }\href
  {https://doi.org/10.1103/PhysRevLett.123.215001} {\bibfield  {journal}
  {\bibinfo  {journal} {Phys. Rev. Lett.}\ }\textbf {\bibinfo {volume} {123}},\
  \bibinfo {pages} {215001} (\bibinfo {year} {2019})}\BibitemShut {NoStop}%
\bibitem [{\citenamefont {Hundhausen}(1995)}]{Hundhausen_Chapter}%
  \BibitemOpen
  \bibfield  {author} {\bibinfo {author} {\bibfnamefont {A.~J.}\ \bibnamefont
  {Hundhausen}},\ }\bibfield  {title} {\enquote {\bibinfo {title} {The solar
  wind},}\ }in\ \href@noop {} {\emph {\bibinfo {booktitle} {Introduction to
  Space Physics}}},\ \bibinfo {editor} {edited by\ \bibinfo {editor}
  {\bibnamefont {{M. G. Kivelson and C. T. Russell}}}}\ (\bibinfo  {publisher}
  {Cambridge University Press},\ \bibinfo {year} {1995})\ Chap.~\bibinfo
  {chapter} {4}, pp.\ \bibinfo {pages} {91--128}\BibitemShut {NoStop}%
\bibitem [{\citenamefont {Mullan}\ and\ \citenamefont
  {Smith}(2006)}]{Mullan_SP2006}%
  \BibitemOpen
  \bibfield  {author} {\bibinfo {author} {\bibfnamefont {D.~J.}\ \bibnamefont
  {Mullan}}\ and\ \bibinfo {author} {\bibfnamefont {C.~W.}\ \bibnamefont
  {Smith}},\ }\bibfield  {title} {\enquote {\bibinfo {title} {Solar wind
  statistics at 1 au: Alfven speed and plasma beta},}\ }\href
  {https://doi.org/10.1007/s11207-006-2077-y} {\bibfield  {journal} {\bibinfo
  {journal} {Solar Physics}\ }\textbf {\bibinfo {volume} {234}},\ \bibinfo
  {pages} {325--338} (\bibinfo {year} {2006})}\BibitemShut {NoStop}%
\end{thebibliography}%

\end{document}